\documentclass[twocolumn]{aastex63}

\newcommand{\teff}{$T_{\mathrm{eff}}$}
\newcommand{\logg}{log $g$}
\newcommand{\loggf}{log $gf$}
\newcommand{\ebv}{$E(B-V)$}
\newcommand{\feh}{[Fe/H]}
\newcommand{\mua}{$\mu_{\alpha}$}
\newcommand{\kms}{km-s$^{-1}$}
\newcommand{\msun}{$M_{\sun}$}
\newcommand{\mud}{$\mu_{\delta}$}
\newcommand{\vrad}{$V_{RAD}$}
\newcommand{\vrot}{$v_{ROT}$}
\newcommand{\vturb}{$\xi$}
\received{date}
\revised{}
\accepted{}
\shorttitle{Fe and Li in NGC 2204}
\shortauthors{Anthony-Twarog, Deliyannis, Steinhauer, Sun, \& Twarog}

\begin{document}

\title{Li EVOLUTION AMONG STARS OF LOW/INTERMEDIATE MASS: THE METAL-DEFICIENT OPEN CLUSTER, NGC 2204} 

\author{Barbara J. Anthony-Twarog}
\affiliation{Department of Physics and Astronomy, University of Kansas, Lawrence, KS 66045-7582, USA}
\email{bjat@ku.edu}
%\and
\author{Constantine P. Deliyannis}
\affiliation{Department of Astronomy, Indiana University, Bloomington, IN 47405-7105, USA}
\email{cdeliyan@indiana.edu}
%\and
\author{Aaron Steinhauer}
\affiliation{Department of Physics and Astronomy, State University of New York, Geneseo, NY 14454, USA}
\email{steinhau@geneseo.edu}
%\and
\author{Qinghui Sun}
\affil{Tsung-Dao Lee Institute, Shanghai Jiao Tong University, Shanghai, 200240, China}
\affil{Department of Astronomy, Tsinghua University, Beijing, 100084, China}
\email{qinghuisun1@gmail.com}
%\and
\author{Bruce A. Twarog}
\affiliation{Department of Physics and Astronomy, University of Kansas, Lawrence, KS 66045-7582, USA}
\email{btwarog@ku.edu}

\begin{abstract}
We have analyzed high-dispersion spectra in the Li 6708 \AA\ region for 167 stars within the anticenter cluster, NGC 2204. From 105 probable members, abundance analysis of 45 evolved stars produces [Fe/H] $= -0.40 \pm 0.12$, [Si/Fe]$ = 0.14 \pm 0.12$, [Ca/Fe] $= 0.29 \pm 0.07$, and [Ni/Fe] $= -0.12 \pm 0.10$, where quoted errors are standard deviations. With $E(B-V) = 0.07$ and $(m-M)_{0}$ = 13.12, appropriate isochrones provide an excellent match from the main sequence through the tip of the giant branch for an age of 1.85 $\pm$ 0.05 Gyr. Li spectrum synthesis produces A(Li) below 1.4 at the base of the red giant branch to a detectable value of -0.4 at the tip. Six probable AGB stars and all but one red clump star have only Li upper limits. A rapidly rotating red giant is identified as a possible Li-rich giant, assuming it is a red clump star. Main sequence turnoff stars have a well-defined A(Li) = 2.83 $\pm$ 0.03 (sem) down to the Li-dip wall at the predicted mass of 1.29 \msun. Despite the same isochronal age as the more metal-rich NGC 2506, the red giant luminosity distribution reflects a younger morphology similar to NGC 7789, possibly indicating a deeper metallicity impact on stellar structure and A(Li) than previously assumed. As in NGC 2506 and NGC 7789, the NGC 2204 turnoff exhibits a broad range of rotation speeds, making abundance estimation impossible for some stars.  The cluster place within Galactic A(Li) evolution is discussed.

\end{abstract}
%\keywords{}

\section{Introduction}

In a pioneering study, \citet{H75a} compiled a catalog of open clusters older than the Hyades using pseudo-color-magnitude diagrams to identify and categorize by age a large sample of previously unstudied open clusters observable from the southern hemisphere, noting a deficiency of older clusters in the direction of the galactic center. For a subset of the newly categorized systems, including three of the oldest, NGC 2243 \citep{H75b}, Mel 66 \citep{H76b}, and NGC 2204 \citep{H76a}, Hawarden combined $UBV$ photographic and photoelectric photometry to constrain the basic properties of each cluster from its two-color and color-magnitude diagrams (CMD). Even with the photometric precision of nearly five decades past, it was readily discernible from the relative positions of the turnoffs and the well-populated giant branches that the clusters were significantly older than the Hyades and, for NGC 2243 and Mel 66, similar to or greater than the canonically old open cluster, M67. For NGC 2204, the sparsely populated subgiant branch indicated an age roughly half that of M67, while $(U-B),(B-V)$ two-color analysis suggested a modest reddening, \ebv $\sim 0.06$ to 0.08 and, critically, a $\delta(U-B)$-based metallicity comparable to Mel 66 but slightly higher than the clearly metal-deficient, anticenter cluster, NGC 2243, less than 15\arcdeg\ south of NGC 2204. Any reasonable estimate of the NGC 2204 distance placed it more than a kiloparsec above the galactic plane, leading \citet{H76a} to classify the cluster as a potential halo, rather than disk, object.

Following an earlier analysis of the open cluster, NGC 2420 \citep{MC74}, Hawarden's cluster program helped initiate a series of studies focused on anticenter clusters of intermediate age and sub-solar metallicities (see, e.g. Mel 66 \citep{AT79}, Ber 21 \citep{CH79}, NGC 2506 \citep{MC81}, and NGC 2158 \citep{CH85}), motivated in part by attempts to use the clusters to define the galactic abundance gradient (e.g. \citet{JA79}). 

Clusters including NGC 2420, NGC 2506, NGC 2204, and NGC 2243 are also studied to provide valuable insight into the role of metallicity on the evolution of stars of intermediate (1.6 \msun) to low (0.9 \msun) mass. From an observational standpoint, NGC 2204 has presented some challenges that have reduced its popularity for programs devoted to anticenter clusters of lower metallicity. It is more distant and significantly less populous than better studied systems like NGC 2420 and NGC 2506, and is neither as old nor as metal-poor as Mel 66 and NGC 2243. Astrometric isolation of members from the rich foreground field has become easier (see, e.g. \citet{DI14, CG18}) only recently, while high precision radial velocity studies of the cluster have reached barely below the level of the red giant clump \citep{ME07}.

Despite these challenges, NGC 2204 was included in our ongoing program to delineate the impact of metallicity, age, and stellar rotation on the evolution of Li among stars of intermediate-to-low mass (see \citet{AT21, SU22} and references therein). As discussed extensively in \citet{DE19}, clusters with ages in the 2 to 4 Gyr range are likely to have main sequence turnoff (MSTO) stars exhibiting Li depletion while unevolved, if their masses placed them in the main sequence Li-dip \citep{BO86}. However, the MSTO stars above (more massive than) the Li-dip may show varying levels of Li depletion, even before cooling and expansion on the subgiant and giant branches lead to depletion of this fragile isotope through dilution due to the deepening of the surface convection zone, and other possible effects \citep{TW20}. As giants, the stars will be low enough in mass ($\leq 2 M_{\sun}$) to experience a disruptive ignition of helium at the tip of the red giant branch (RGB).  Despite the general diminution of Li between the MSTO and giant branch tip, a small number of cluster giants show measurable, and in some cases abundant, levels of surface Li (see e.g. \citet{CA16}).  Since the mass range of the Li-dip is particularly sensitive to metallicity \citep{CU12, AT21}, the combination of cluster age and lower metallicity for NGC 2204 implied that the stars leaving the main sequence in this cluster would be on the hot side of the Li-dip and potentially could still retain the signature of their primordial cluster Li abundance, thereby supplying a constraint for both stellar and galactic chemical evolution.

The goal of the current investigation is to present the results of a spectroscopic survey of NGC 2204 stars from the tip of its extended giant branch to the main sequence turnoff, reaching to the level of the Li-dip.  Section 2 details the experimental design and data acquisition for our spectroscopic observations. Section 3 updates the status of cluster membership and possible binarity for the stars of interest while Section 4 reviews the evidence from prior studies pertaining to the cluster's age, reddening and metal abundance.  Section 5 presents the metallicity determination from the Hydra spectra, followed in Section 6 by the determination of the Li abundances and their contribution to our understanding of stellar and galactic Li. Section 7 contains a summary of our conclusions.

In the following discussions and tables, 
individual stars will be referenced by their WOCS (WIYN Open Cluster Survey) numbers, assigned in a forthcoming photometric survey by \citet{con} (SD24). 
Where available, identifications from WEBDA will also be included with a ``W'' prefix; these identifications are in most cases taken from \citet{H76a} and are the most common identifier in earlier studies of the cluster.

\section{Experimental design: sample selection and data acquisition}
\subsection{Original Sample Selection}

Our spectroscopic sample was constructed in 2014, so without the insight subsequently provided by $Gaia$ astrometric, kinematic and photometric data (\citet{GA16}, \citet{GA18} DR2, \citet{GA22} DR3). A primary goal was the delineation of Li abundance as stars evolve from the hot side of the Li-dip to the base of the giant branch, as exemplified by the analysis of NGC 2506  \citep{AT18b}. 

The extensive radial velocity survey of \citet{ME07} provided an important guide for choosing candidate giant members, as their original discrimination between single members and binary and/or nonmember stars was excellent. Our MSTO candidate list was chosen based on a limited set of instrumental extended Str{\"o}mgren photometry obtained at the same time as the photometric survey of NGC 2506 \citep{AT16}. For the subgiant branch, however, photometric isolation of likely members proved difficult, so all stars within a plausible ($V$, $(B-V)$) range were observed in the hope of identifying a handful of possible subgiants.

With the hindsight that $Gaia$ results have provided, it is made clear in Section 3 that this approach worked well to select stars with joint photometric similarities of temperature and metallicity along the nearly vertical turnoff, but confirmed the minimal presence of member stars within the Hertzsprung gap.

\subsection{Hydra Data Acquisition}
Spectra for 167 stars in the field of NGC 2204 were obtained in 2014 and 2015 using the Hydra multi-object spectrograph on the WIYN 3.5-meter telescope. To cover different magnitude ranges of candidate stars, four separate fiber configurations were developed, the brightest of which incorporated only two stars with $V \leq 12$. This configuration was observed in two exposures on 26 January, 2014 (UT date) for 30 minutes total.  A second configuration covered 45 additional giants to $V = 14$ and was observed in three exposures for 2.6 hours on 25 February, 2014. Two additional configurations were designed for fainter stars and required observations in February 2014 and January 2015. A configuration incorporating 61 stars, largely subgiant candidates, was observed for nearly eight hours on 25 January 2014 and 20 February 2015, with the final and faintest configuration of 59 MSTO candidates receiving over sixteen hours of exposure on 27 February 2014, 18 and 19 January 2015. We note that our fiber configurations also incorporated dozens of unassigned fibers from which simultaneous spectra can be used for sky subtraction. 

The adopted spectrograph setup produces spectra centered on 6650 \AA\ with a dispersion of 0.2 \AA\ per pixel and a range of $\sim400$ \AA. Examination of thorium-argon lamp spectra, used for wavelength calibration, indicates lines 2.5 pixels wide, yielding an effective spectral resolution of 13,300. In addition to longer Th-Ar lamp spectra obtained during the day, comparison lamp spectra were obtained before and after object exposures in the course of the night.  Except for radial velocity standards observed throughout the night, comps, dome flats and day-time sky spectra were obtained with the same fiber configurations used for program observations.  These daytime solar spectra were used to correct for fiber-to-fiber throughput differences, as well as provide reference solar spectra to zero the \loggf\ values of individual lines required to reproduce solar abundances. 

Our IRAF\footnote{IRAF is distributed by the National Optical Astronomy Observatory, which is operated by the Association of Universities for Research in Astronomy, Inc., under cooperative agreement with the National Science Foundation.}-based processing steps have been described in past papers (see, e.g. \citet{AT18a, AT18b, DE19, AT21}) and include the typical application of bias subtraction, flat-fielding using dome flats for each configuration, and wavelength calibration using comparison lamp exposures. Our strategy for cosmic ray cleaning uses ``L. A. Cosmic''\footnote{http://www.astro.yale.edu/dokkum/lacosmic/, an IRAF script developed by P. van Dokkum (van Dokkum 2001); spectroscopic version.} on the long exposure frames after the flat field division step. Final composite spectra were obtained by co-addition of multiple exposures to obtain the highest possible Signal-to-Noise per pixel ratio (SNR). The combination of spectra from 2014 and 2015 was completed only after inspection for possible radial velocity variability. One star has a noticeable wavelength shift over the year-long interval and will be noted in a subsequent table with separate velocity estimates based on individual rather than summed datasets.

Since the goal of co-addition is to obtain higher SNR spectra, it is worth noting how that parameter may be estimated and what the resulting values mean.  One criterion estimates the total flux-per-pixel above sky before continuum fitting, an indication of the total signal accumulated for a star.  The square root of this quantity provides a fair estimate of the accumulated flux-based SNR.  Another method measures the actual variance of the spectrum in a line-free region.  Even in a relatively metal-poor star, such a region is hard to come by and tends to be relatively small; in the current case, the region between 6680 and 6690 \AA\ was selected for this purpose. For the cooler giant stars these latter SNR estimates, derived using $splot$ in IRAF, are significantly smaller than the flux-based estimators by factors as large as dozens, primarily because numerous difficult to separate weak lines begin to appear, making a line-free region difficult to identify. The flux-based SNR will be adopted in all further discussions.
Other than one MSTO star with SNR $=83$ due to a single epoch of observation, all stars had SNR in excess of 100.  A few luminous RG have coadded spectra with SNR over 700.  For the 105 members, the median SNR value is 217.  

The Fourier-transform, cross-correlation utility {\it fxcor} in IRAF was used to assess kinematic information for each star from the summed composite spectra. In {\it fxcor}, program stars are compared to zero-velocity stellar templates of similar $T_{\rm{eff}}$.  Templates are usable for spectral types F5 through K6, making this technique less effective for the reddest members of NGC 2204. The {\it fxcor} utility characterizes the cross-correlation-function (CCF), from which estimates of each star's radial velocity are easily inferred. Formal errors implied by the CCF analysis with $fxcor$ suggest that typical formal \vrad\ errors are $\sim 1$ \kms.  We were in a position to test this more formally as two stars were observed in duplicate configurations straddling the two years of our survey.  Discrepancies within those year-to-year and configuration-to-configuration comparisons were well under 1 \kms\ for the two stars. Radial velocity standards were observed every night of the 2014/2015 observing runs, ideally in the same configuration and time-adjacent to the program observations. In all, observations of six standards on five separate nights generated the 16 individual radial velocity measures used to zero the {\it fxcor}-derived radial velocities to the standard system to within 1.0 \kms.

Rotational velocities can also be estimated from the cross-correlation-function full-width-half-maximum (CCF FWHM) using a procedure developed by \citet{ST03}. We note here that although we will refer to rotational velocities, the measurements in fact yield projected rotational velocities $V_{sin i}$. Our method exploits the relationship between the CCF FWHM, line widths and \vrot, using a set of numerically ``spun up" standard spectra with comparable spectral types to constrain the relationship. Our previous analyses of both red giant and turnoff spectra demonstrate that $fxcor$ cross-correlation profiles have significantly reduced accuracy when attempting to reproduce rotational velocities above 35 \kms.  We note that spectral resolution alone implies that derived projected rotational velocities below $\sim$15 \kms are not meaningful, a result confirmed in past analyses (see, e.g. \citet{DE19}). With the exception of stars likely to be binaries, the derived mean projected rotational velocities are unexceptional:  $14.7 \pm 3.5$ \kms  for the RG and SG members not identified as photometric or \vrad\ variables in other surveys, implying that, as expected for evolved stars, the stars have spun down to minimal rotation speeds.  MSTO stars not designated as potential binary or wide-lined show an average \vrot=$25.5 \pm 10.5$ \kms.  By contrast, MSTO stars flagged as potential binaries and/or wide-lined based on visual examination of spectra show predictably larger average \vrot\  values of $34.3 \pm 15.4$ \kms.

\subsection{External Sources of Spectra}
We close this section by noting additional spectra for four giant members of NGC 2204, accessed from the ESO Science Archive Facility for comparison and analysis.  These spectra had been obtained for program ID 167.D-0173, PI R. Gratton in October, 2001.  The spectra, with SNR ranging from 116 to 179, were obtained with the Fibre Large Array Multi Element Spectograph (FLAMES) fiber-feed assembly to the high resolution UVES spectrograph with pixel-resolutions of 16.9 m\AA.

\section{Cluster Membership and Binarity}
\subsection{Astrometry} 

For membership purposes, the first phase of estimation comes from the astrometric contribution supplied by the proper motion and parallax, exquisitely measured by the ongoing $Gaia$ collaboration. The first comprehensive analysis of the $Gaia$ data for clusters was compiled by 
\citet{CG18} (CG18), with periodic updates since then \citep{CG20} (CG20).

While all nonzero membership probabilities for stars within NGC 2204 were available from CG18, to protect against omission of potential members at large radial distance from the cluster center or those tagged as 0 probability members due to other anomalies such as poor astrometric measure, we adopted a process of membership discrimination similar to that used in the much richer but less distant cluster, M67 \citep{TW23}. As a baseline, we made use of the cluster membership survey by CG18, tied to the DR2 data release.  Necessary first steps involved cross-matching of $BV$ photometric indices from SD24 with cluster membership lists from CG18 and the raw data from $Gaia$ DR3. For a few spectroscopic candidates which fall outside the SD24 field of study, $Gaia$ synthetic photometric values ($GSPC$) \citep{gspc} for $V$ and $B$ have been used. 

It was initially assumed that absence of a star from the original CG18 member compilation implied nonmembership though, as noted, the possibility existed that stars with larger than typical astrometric errors could be incorrectly excluded from the cluster database.  Since astrometric data for all sample stars are readily accessible in the DR3 data, ``nonmembership'' for stars not tagged as members by CG18 has been quantified as follows. 572 stars in NGC 2204 classified as members with a probability higher than 50\%  (CG18) were identified in DR3. Using only 258 stars from DR3 with $\pi / \sigma_{\pi} \geq 5$, mean cluster values in $\pi$ and proper motion were derived, generating $\pi$ = 0.224 $\pm$ 0.041 mas, \mua\ = -0.575 $\pm$ 0.081 mas-yr$^{-1}$, and \mud\  = 1.958 $\pm$ 0.072 mas-yr$^{-1}$, where the quoted errors refer to the standard deviations about the mean. For each observed candidate star in DR3, a quantity QM (quality metric) was constructed from the square root of the quadratic sum of the number of standard deviations that star's $\pi$, \mua\ and \mud\ are from the mean values.  Stars classified as members from the CG18 compilation have QM of 10 or less within DR3, {\it i.e.}, probable member stars are less than $10 \sigma$ in all three astrometric dimensions from the cluster mean parameter values. Stars in our target sample that are not included in either CG18 or CG20, and are thus likely nonmembers, have QM values from 11.5 to as high as 2000. 

\begin{figure}
\figurenum{1}
\includegraphics[angle=0,width=\linewidth]{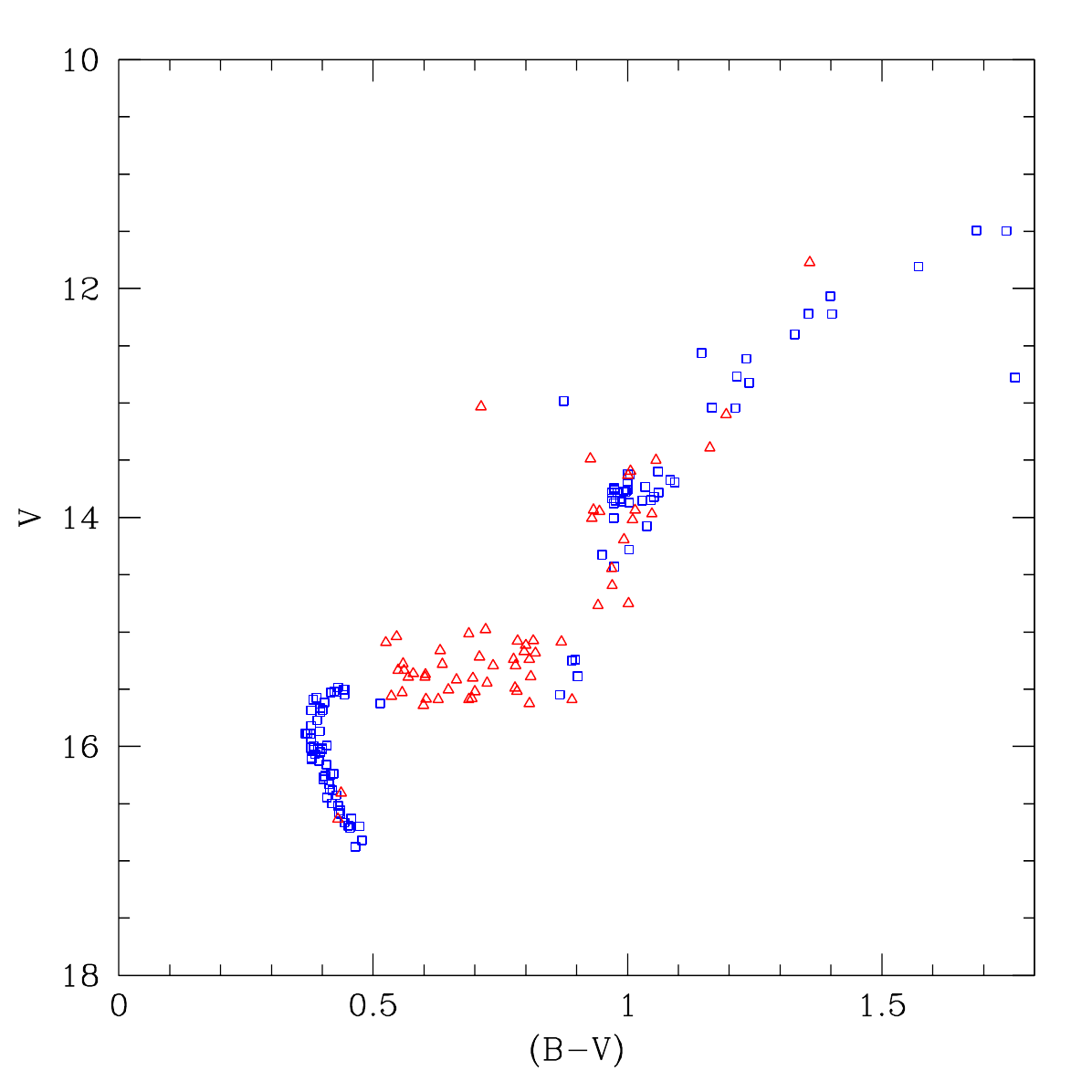}
\caption{CMD of the preselected sample of 167 stars in NGC 2204.  Ultimate membership decisions are reflected by symbol: blue squares for probable members, red triangles for nonmembers.}
\end{figure}

Figure 1 illustrates the CMD location of the Hydra sample stars; symbol colors and shapes reflect the subsequent separation into members and nonmembers, based purely on astrometric criteria. Our original photometrically-based member selection was extremely successful for the MSTO region, although it necessarily short-changed the more interesting aspects of the blue hook morphology associated with the rapid hydrogen exhaustion phase (see the isochrones of Figure 4 between $V$ = 14.8 and 15.2) to maintain a sample of easily identifiable stars on the vertical turnoff.  With decades of prior radial velocity observations to guide our sample selection of giants, that configuration, too, was largely successful in tagging member stars as targets.  Our attempt to study the meager subgiant branch, however, was primarily unsuccessful. As will be discussed in Section 6, this deficiency of  stars on the horizontal subgiant branch is real and is a strong indicator of the relative youth of NGC 2204 compared to the 3.6 and 3.7 Gyr-old clusters, NGC 2243 \citep{AT21} and M67 \citep{TW23}. From the more complete astrometric sample of Figure 4 and excluding stars that lie well off the isochrone sequence, 8 stars are positioned between the MSTO and the RGC; our spectroscopic sample includes 5 of these, primarily those at the base of the vertical FRG branch.

\subsection{Radial Velocity Determination: Current and Past}

Results based on $fxcor$ for the MSTO stars in a metal-poor cluster like NGC 2204 can be challenging. Several contributing factors such as rotation and/or binarity might conspire to broaden and blur the already relatively weak lines. Although our spectra are of more than adequate SNR per pixel, $fxcor$ struggled with weak and apparently washed out lines. For many such MSTO stars, velocities are reported, but the larger than usual formal errors indicate where the results are less reliable. 

Although not directly used as a membership discriminant, our radial velocity data consistently support the astrometric separation into members and nonmembers.  For the 105 stars ultimately classed as astrometric members, the mean radial velocity is $92.5 \pm 11.3$ \kms; the larger than average dispersion is likely due to the contribution of binaries which have not been identified and eliminated from the sample, as well as the uncertainty added by rapid rotators with broadened lines. The velocities for the designated nonmembers are gratifyingly distinct:  $41.75 \pm 40.0$ \kms. For our best estimate of the cluster radial velocity, we can eliminate 6 stars (see discussion below) previously identified as spectroscopic binaries, plus the star WOCS5015 which, while an astrometric member, has a radial velocity, confirmed by Gaia, near 21 \kms. For the 99 remaining stars, the mean radial velocity becomes 92.45 $\pm$ 8.18 \kms, still a disturbingly large scatter. The source of the problem becomes obvious when the sample is reduced to only red giants and subgiants, stars with intrinsically narrow lines. For these 42 stars, the cluster mean becomes $90.9 \pm 1.94$ \kms. The fact that the dominant source of the scatter in the original cluster mean comes from broad-lined stars at the turnoff is confirmed by the direct comparison between our individual radial velocities and those from past research, usually on evolved stars, where the scatter in the offset residuals is typically 0.7 - 2.8 \kms (see discussion below).  

Although still quite important in the context of membership discrimination, radial velocity plays a growing role in the identification of potential variable and binary systems by highlighting departures from cluster mean properties. A review of past radial velocity surveys will set up a discussion of our survey results and comparisons.
 
\citet{JA11} summarize the various attempts made before 2000 to define the cluster mean radial velocity, including \citet{FR02} who observed 19 red giant candidates, 7 of which were classified as nonmembers or possible nonmembers. \citet{FR02} found a mean cluster velocity of $89 \pm 6$ \kms. 

\citet{ME07} published velocities for giants in several open clusters based on observations with the CORAVEL instrument at the Danish 1.5-m telescope at La Silla, Chile, including 35 stars in NGC 2204.  Ten stars were identified as nonmembers, with an additional handful of stars flagged as spectroscopic binary candidates, several of which will be discussed in greater detail below. From the single-star candidate members, a mean radial velocity of $91.38$ \kms\ was obtained, with an rms error of 1.33 \kms. \citet{ME07} noted the agreement within the errors with the earlier, less precise result of \citet{FR02}.

\citet{JA11} conducted a Hydra II-based survey of 35 stars in NGC 2204, obtained at the 4-m Blanco telescope at CTIO.  
Based on 16 stars, they arrived a mean radial velocity of $88.4 \pm 1.3$ \kms.  An additional ground-based spectroscopic survey of interest is that of \citet{CA16}, supplying radial velocities for NGC 2204 based on spectra  obtained with the MIKE instrument at Magellan with R $\sim 31,000$ to 44,000. From an original dataset of 19 candidate evolved stars, \citet{CA16} excluded two stars, WOCS1003=W4119 and WOCS8011=W2330, as a potential binary and nonmember respectively to arrive at an average radial velocity $91.12 \pm 1.22$ \kms, where the quoted error describes the standard deviation for the set. 

Finally, the recent publication of $Gaia$ DR3 data provides a large set of well characterized radial velocities to which our Hydra values are compared in Figure 2 for 64 stars in common.  Although the most recent data release does not explicitly flag stars for radial velocity variability, candidate binaries might be identified when compared with data from other sources and epochs. Some of the features of Figure 2 refer to stars flagged as potential SB systems in earlier studies or by the present work.

In summary, eliminating seven stars from our sample that exhibit evidence for \vrad\  variability (see Section 3.3 below), comparison of our Hydra \vrad\  results with past surveys produces the following systemic differences. With respect to \citet{FR02}, our results for 11 stars in common are $1.57 \pm 5.65$ \kms\ larger; based on 18 stars in common with \citet{JA11}, our \vrad\ values are $2.80 \pm 2.73$ \kms\ larger.  With respect to \citet{ME07}, the difference in the same sense of (HYDRA - survey) is $-0.26 \pm 2.11$ \kms, based on 20 stars in common.  Finally, comparing to values for 17 stars from \citet{CA16}, the difference is $-0.11 \pm 0.74$ \kms. Even with the exclusion of seven potentially variable velocity stars, 57 stars remain in common between the Hydra sample and results from DR3 for which the average difference, (HYDRA - DR3) is $-1.03 \pm 1.69$ \kms.  Typical velocity errors for these datasets are $0.92 \pm 0.25$ \kms\ for Hydra values, $1.73 \pm 1.10$ for DR3 values.

\subsection{Radial Velocity Variables and Binarity}
\citet{ME07} identified one star as a clear spectroscopic binary (SB), WOCS1002=W1129, identified in Figure 2 by a large red filled circle.  In spite of a high probability of membership in CG18,  its $Gaia$ DR3 velocity is offset both from the cluster mean and the Hydra value by more than 6 \kms.  Four other stars (WOCS1005=W4132, WOCS2006=W1136, WOCS4014=W3304, WOCS1003=W4119) were flagged by \citet{ME07} as potential spectroscopic binaries and are noted in Figure 2 with open magenta triangles.  WOCS1005 is a very cool giant near the bending tip of the RGB.  Identified as an L non-periodic variable by ASAS-SN \citep{SH14}\footnote{https://asas-sn.osu.edu/variables} the star is also flagged as a variable in DR3.  Our Hydra-based radial velocity was based only on the H$\alpha$ line, resulting in a large formal error for that value:  $104.5 \pm 8.9$ \kms\, compared with a DR3 value of $94.3 \pm 1.16$ \kms. Like WOCS1005, WOCS2006 also has an extremely red $(B-V)$ color, is flagged as a photometric variable in DR3 and has been classified by ASAS-SN as a semi-regular variable with a 59.4-day period. The remaining two stars flagged by \citet{ME07} as possible SBs inhabit less extreme locations along the RGB.  WOCS4014 and WOCS1003 were both noted as a potential SB by \citet{JA11}. As mentioned above, WOCS1003 was also identified as a spectroscopic binary by \citet{CA16}.  Although only observed in 2014 in our Hydra survey, our result is consistent with a spectroscopic binary classification for the star. The 2014 \vrad $= 129.5 \pm 0.9$ \kms\ may be compared to the $Gaia$ \citep{GA18} (DR2) value of $80.0 \pm 1.5$ \kms\ and the DR3 result of $87.9 \pm 4.3$ \kms, supporting a case for \vrad\ variability for this star.  The large error for the DR3 results may in itself be indicative of variability within the composite sets of data comprising this final value.

Returning to the case of WOCS8011=W2330, noted by \citet{CA16} as a potential radial velocity nonmember with a \vrad\  of 96.5 \kms, we note some evidence of radial velocity variability in our results and those culled from DR3.  Our Hydra value of $108.8 \pm 0.8$ \kms\  is similar to the DR3 value of $101.6 \pm 4.4$ \kms\ but even more widely separated from the cluster mean.  We identify this star as a possible SB.

\begin{figure}
\figurenum{2}
\includegraphics[angle=0,width=\linewidth]{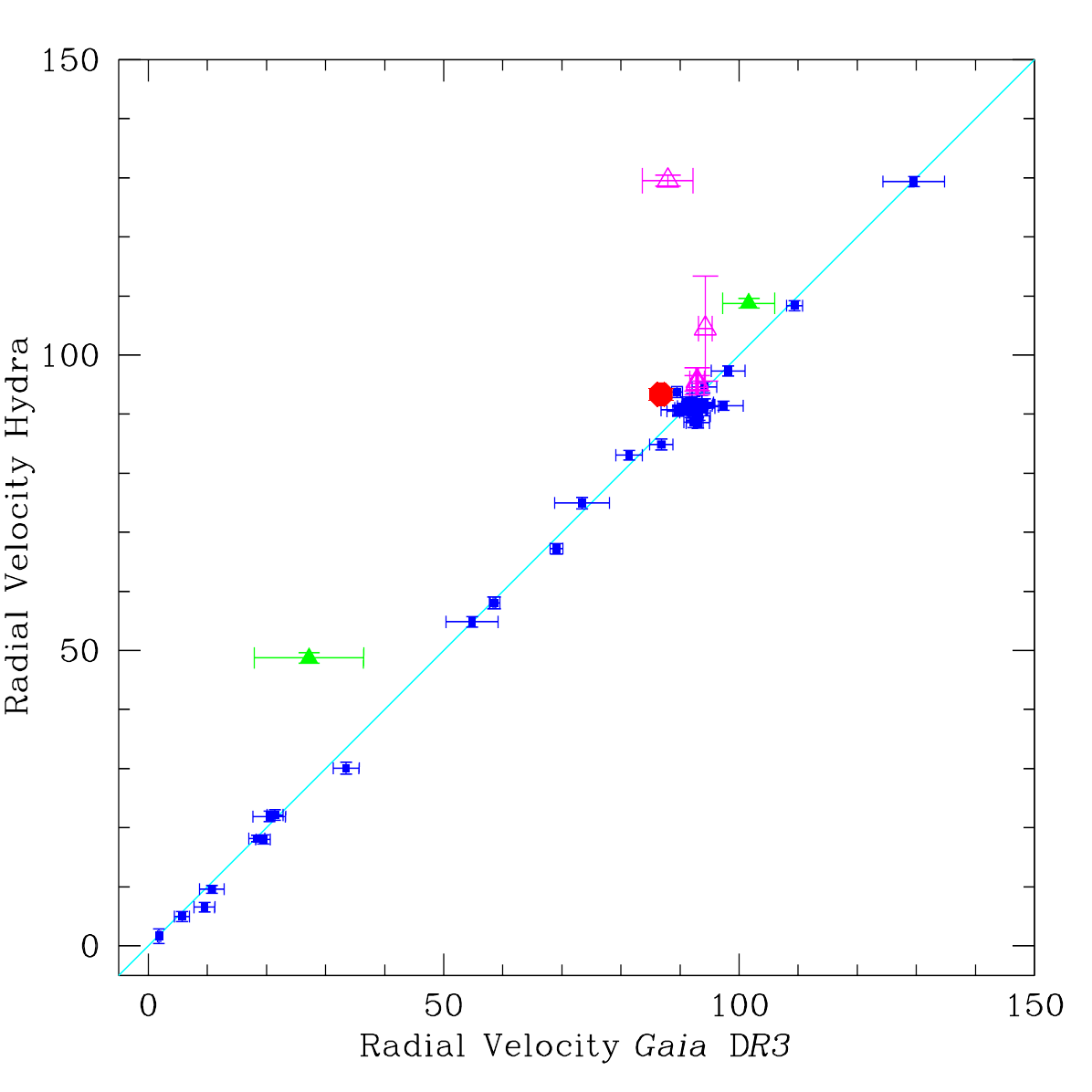}
\caption{Comparison of Hydra and $Gaia$ DR3 radial velocities. The cyan line is not a fit, merely an illustration of one-to-one correspondence. A single red filled symbol shows WOCS1002, denoted by MM as an SB.  Possible SBs WOCS1005, WOCS2006, WOCS1003 and WOCS4014 are denoted by magenta open triangles.  Two green solid triangles show the positions of WOCS9015 and WOCS8011 while the blue squares represent the remaining comparisons between Hydra and DR3 radial velocities.}
\end{figure}

Although not found in any other \vrad\ survey, results for WOCS11733 indicate it to be a nonmember and potential radial velocity variable. The separate Hydra observations in 2014 ($44.3 \pm 1.5$ \kms) and 2015 ($83.3 \pm 0.7$ \kms) imply binarity.  This star will not be considered further, as both astrometry and the \vrad\ values indicate nonmembership. 
One more radial velocity variable candidate is found among the apparent nonmembers.  Star WOCS9015=W1308 appears among the fainter RGB stars but has radial velocity measures that suggest nonmembership and potential binarity;  the $Gaia$ DR3 recorded value is 48.73 \kms with a typical error.  Our Hydra measurement is considerably smaller, 27.2 \kms, but with an atypically large error of 9.24 \kms.  Astrometry from $Gaia$, discussed below, supports nonmembership for the star.  These two stars are represented in Figure 2 by filled green triangles.

\subsection{Indications of Variability or Binarity from Photometry}
Many open clusters have been closely observed for main sequence variable candidates by J. Kaluzny and collaborators, including NGC 2204 \citep{kal}. Six variable candidates were identified in the cluster, of which two are bright enough to have been included in the present study.  The group's temporal coverage was sufficient to establish light curves and classifications for all six candidates. The brighter candidate (their number 438, V0402 CMa, WOCS5002) is classed as a detached, circularized eclipsing binary with a period of nearly seven days. The other candidate (226, V0403 CMa, WOCS20010, W2216)  may be an ellipsoidal variable with a period $\geq$ 2d. Both stars are highly probable members according to CG18 and one of the two (WOCS20010) is in our Hydra sample.

The ASAS-SN database was also probed for variables, identifying nine candidates with $V \leq 17$ and within 20\arcmin\  of the center of NGC 2204.  Of the nine, four are not in the CG18 membership list; two others are identical to the two variable candidates identified by \citet{kal}.  The remaining three are evolved stars, including two of the reddest stars among the giants of NGC 2204.  WOCS1005 = W4132 and WOCS2006=W1136 have $(B-V)$ colors redder than 1.7.  WOCS1005 is not in the CG20 member list but its astrometric properties suggest that it is a possible member. ASAS-SN describes it as nonperiodic L type variable.   WOCS2006 is a likely member with a period of 59 days and was noted initially by \citet{H76a} as a semi-regular variable distinctive for its position below the main giant branch. 

The final ASAS-SN designated variable candidate is WOCS4009=W4216, denoted as a rotational variable, i.e. a star with variable brightness probably due to its rotation coupled with a non-uniform surface brightness, with a period of 6.18 days. Its astrometric parameters indicate that it is a member with a CMD location near the RGB clump. Spectra for the four giants observed by VLT are shown in Figure 3,  where the unusual breadth of WOCS4009 is easily noted, as is its velocity offset of $\sim30$ \kms\ from the other three giants.

\begin{figure}
\figurenum{3}
\includegraphics[angle=0,width=\linewidth]{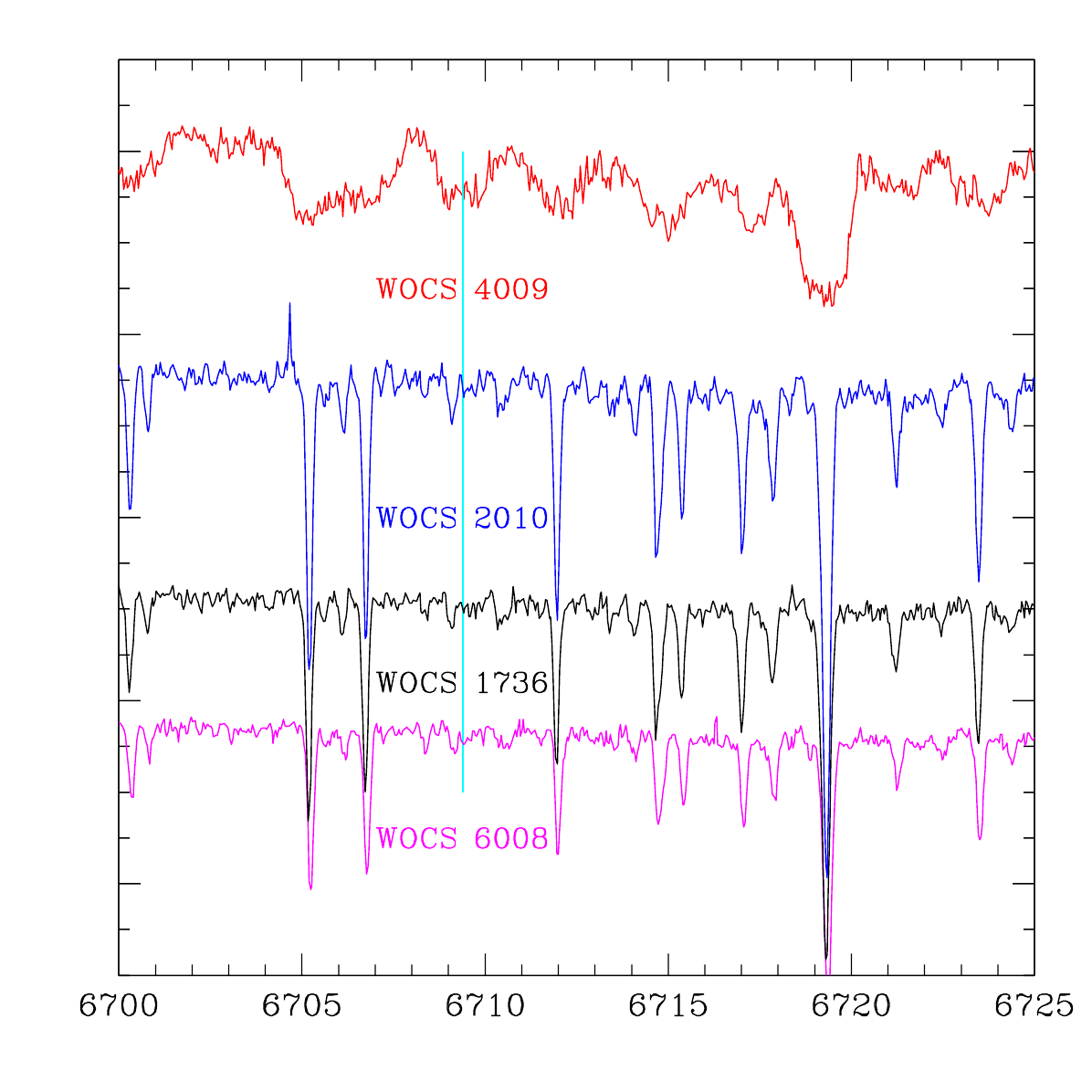}
\caption{VLT spectra in the Li line region for four giant members of NGC 2204, including the apparently rapid rotator W4216=WOCS4009.  For this figure, the spectrum of WOCS4009 has been offset by 0.7 \AA\ to the blue, equivalent to a $\sim 30$ \kms offset to align its spectrum with the other three.  The vertical cyan line indicates the expected location of the Li line, 9.9 \AA\ to the blue of the Ca 6717 \AA\ line.}
\end{figure}

Among the $Gaia$ data products released in DR3 is a photometric variability flag.  The three giants included among the $Gaia$ variable candidates are the same stars similarly flagged by ASAS-SN. An additional half dozen MSTO members are among the stars flagged for variability, including the star identified by \citep{kal} as an ellipsoidal variable, WOCS20010.  

\floattable
\begin{deluxetable}{rrhccrrrrrc}
\tablenum{1}
\tablecaption{Hydra Sample}
%\tabletypesize{small}
\tablewidth{0pt}
\tablehead{
\colhead{$\alpha(\rm{DR3})$} & \colhead{$\delta(\rm{DR3})$} & \nocolhead{ConNum} & \colhead{WOCS} & \colhead{WEBDA} &
\colhead{$V$} & \colhead{$(B-V)$} & \colhead{Status} & \colhead{$V_{RAD}$} & \colhead{$\sigma_{Vrad}$}  &
\colhead{Notes} } 
\startdata
93.90276 & -18.78131 & 2541 & 2014 & 3325 & 11.491 & 1.686 & M & 91.22  & 2.33 & 8 \\
93.88815 & -18.70460 & 2368 & 1005 & 4132 & 11.495 & 1.745 & M & 104.48 & 8.87 & 4:A,G; 7:H, MM-SB?; 8; 9 (Note a)\\
94.08091 & -18.83833 & 4220 & 1732 &      & 11.771 & 1.358 & N & 1.65  & 1.25  &  \\
93.88085 & -18.71567 & 2281 & 1006 & 4137 & 11.806 & 1.572 & M & 89.31 & 1.71  & 4\\
93.74355 & -18.64528 & 897 & 1017  &      & 12.065 & 1.399 & M & 91.36 & 1.06  &  \\
93.70613 & -18.52075 & 589 & 1028  &      & 12.220 & 1.356 & M & 91.81 & 1.02  & 9 \\
93.99237 & -18.67401 & 3467 & 4014 & 3304 & 12.221 & 1.402 & M & 95.24 & 1.30  & 6; 7:MM-SB?\\
94.08265 & -18.77434 & 4231 & 2028 &      & 12.400 & 1.329 & M & 91.18 & 1.00  & 9\\
93.83926 & -18.59759 & 1792 & 1011 & 1320 & 12.562 & 1.146 & M & 90.97 & 0.88  & 8\\
93.88300 & -18.66028 & 2309 & 1002 & 1129 & 12.613 & 1.234 & M & 93.36 & 1.03  & 7:MM- SB\\
93.95708 & -18.62761 & 3125 & 3011 & 2212 & 12.766 & 1.215 & M & 91.39 & 0.95  & 6\\
93.87297 & -18.62536 & 2182 & 2006 & 1136 & 12.777 & 1.762 & M & 95.57 & 2.29  & 4:A,G,H; 7:MM-SB?;8; 9\\
93.91828 & -18.77696 & 2716 & 6014 & 3324 & 12.822 & 1.239 & M & 90.84 & 0.98  &  \\
93.94558 & -18.63205 & 3004 & 2009 & 2211 & 12.982 & 0.875 & M & 91.52 & 0.85  & 5; 8\\
93.91942 & -18.76232 & 2728 & 1012 &      & 13.032 & 0.712 & N & 18.16 & 0.53  & 4:G; 8;9\\
\enddata
\tablecomments{Code for notes: (1) Star is outside photometric survey area, $V$ and $B-V$ synthesized from $Gaia$ DR3 $GSPC$; (2) Only VLT spectrum; (3) Also VLT spectrum; (4) Photometric variability designated by $Gaia$ (G), Hawarden (H), \citet{kal} (K) or ASAS-SN (A) (5) DR3 indicates non-single star; (6) Reported radial velocity is an average from two epochs and/or configurations -- reported error is average of each epoch's formal error; (7) Radial velocity variation detected from Hydra spectra or reported by \citet{ME07} (MM); (8) $Gaia$ magnitude $g$ and $V$ appear discrepant; (9) $Gaia$ $ruwe$ significantly different from 1.0; (10) no astrometry in DR3.}
\tablecomments{Individual notes: (a) Hydra radial velocity estimated from H$\alpha$ only; (b) \vrad\ variability suggested by comparison to DR2 ($79.96 \pm 1.34$ \kms), DR3 ($87.94 \pm 4.29$\kms);(c) \vrad\ variability for WOCS4009 suggested by VLT offset and DR3 value; (d) \vrad\ values for WOCS11733 in 2014 and 2015 are respectively $44.34 \pm 1.46$ \kms\ and $83.3	\pm 0.74$ \kms; (e) WOCS24009 reports single epoch only.} 
\end{deluxetable}

Table 1 contains a summary of our basic information for the 167 stars included in the spectroscopic survey, comprising 105 members and 62 likely nonmembers.
The columns in Table 1 include the IDs, if available, from WEBDA (\citet{H76a}) as well as WOCS numbers. A handful of stars are outside the area covered by SD24 and are noted by letters A through E in the complete Table 1. Also listed are ($\alpha$, $\delta$) for each star from DR3, $V$ and $(B-V)$ from the broad-band photometric survey by SD24 or synthesized photometry from $Gaia$, our derived radial velocity with errors and our assessment of membership status. 

\section{Preliminary Cluster Properties: Reddening, Metallicity, Age}

The analysis and interpretation of our spectra will require knowledge or assumptions about the cluster's bulk properties of foreground reddening, \ebv, metallicity and age so that atmospheric parameters needed for model atmosphere construction may be matched to the stars' \teff\ and \logg\ values, the latter estimated by comparions of photometric values to well-matched isochrones.  Precise, extensive and deep photometry is therefore necessary.  We utilize photometric colors to estimate \teff\ using a color-temperature relation also dependent on \ebv\ and \feh\ estimates. Contributions from past studies that are most relevant to determinations of \ebv\  and \feh\  are summarized below. Where appropriate, conclusions from these prior studies have been reevaluated in the light of astrometric membership information.

\citet{H76a} provided early evidence for NGC 2204's status among metal-poor, lightly reddened and intermediate-age open clusters.  More than two decades would pass until deeper and wider-field photometric surveys would supersede that work but evidence for the modest foreground reddening and metal paucity of the cluster continued to reinforce Hawarden's basic conclusions, including a photoelectric $DDO$ study by \citet{D81}.  For six of the stars observed by \citet{D81}, internally consistent photoelectric techniques permitted a reddening estimate of \ebv$ = 0.08 \pm 0.01$.  Reanalysis of the DDO sample using 5 probable members by \citet{TAAT} led to \ebv = 0.08 and [Fe/H] = -0.34 $\pm$ 0.25 (sd).

In the hindsight enabled by $Gaia$, several of the stars used by Dawson are not found in the cluster membership list assembled by CG18. Inspection of the derived reddening and $\delta CN$ parameters for six probable members as derived in the current analysis suggests an average \ebv $= 0.04 \pm 0.07$ and [Fe/H] = $-0.53 \pm 0.34$, citing simple standard deviations among the separate stars' estimates.  The $DDO$ estimator of metallicity is based on $\delta$CN and thus might be better viewed as an estimate of [m/H] rather than a purely iron-peak abundance estimate but, given the large dispersion, is of little fundamental value.

A deep and extensive $BVI$ CCD survey of NGC 2204 was published by \citet{K97} which incorporated comparisons of $BV$ and $VI$ CMDs to isochrones from \citet{Bertelli94}. While independent determinations of reddening and metallicity were not attempted, some exploration of the interdependence of age, metallicity and reddening was presented for sub-solar metallicities, ages between 1.3 and 2.5 Gyr and reddening values below \ebv = 0.2. Given that the isochrones used have been superceded by more recent models, the absolute values obtained are of questionable value.

It should be noted that independent estimates for \ebv, {\it i.e.} those which don't depend on photometry or spectroscopy,  are available based on dust emission/reddening maps by \citep{SFD, SF11}. The maximum reddening in the direction of NGC 2204 is \ebv = 0.08 to 0.09 based on \citet{SF11}. 

In spectroscopic surveys aimed at abundance determination, reddening-corrected photometric colors are often used with an assumed value for \ebv\ and a color-temperature calibration to provide values or starting estimates for \teff.  Such was the case for \citet{JA11} who adopted a reddening value \ebv $=0.08$ (essentially unchanged from \citet{H76a}) to derive [Fe/H]  $= -0.23 \pm 0.04$ from 13 cluster members. This is higher than but within the errors of the previous result from \citet{FR02}, [Fe/H] = $-0.32 \pm 0.10$ based on somewhat lower resolution spectra.  Spectra for both studies were obtained at CTIO, with the 2011 data obtained using the southern analog of the Hydra instrument used for the present study and in the same wavelength region.

SD24 compiled {\it UBVRI} photometry for over 3800 stars in the field of NGC 2204, centered on ($\alpha,\delta = 93.882$\arcdeg, $-18.650$\arcdeg) using the Half-Degree Imager (HDI) at the WIYN\footnote{The WIYN Observatory was a joint facility of the University of Wisconsin-Madison, Indiana University, Yale University, and the National Optical Astronomy Observatory.} 
0.9-m telescope on Kitt Peak.  Their analysis of the CMD and color-color diagrams led to estimates for NGC 2204's age of $2.2 \pm 0.1$ Gyr, foreground reddening \ebv $= 0.08 \pm 0.01$ and [Fe/H] $= -0.45 \pm 0.05$, based on comparison to Yale-Yonsei \citep{DE04} isochrones.  This survey provides the base $BV$ photometry used in our analysis to develop atmospheric parameter estimates. For a small subset of stars with spectroscopy that lie outside the SD24 survey field, extensive synthetic photometry is now available as part of the DR3 release of $Gaia$ data products \citep{gspc}. We constructed a preliminary comparison between $BV$ data from SD24 and the synthetic $GSPC$ ($Gaia$ Synthetic Photometry Catalog) photometry.  We found from 230 stars with $V \leq 15$ that the SD24 $(B-V)$ colors are $0.009 \pm 0.031$ redder than the $GSPC$ colors, an unconcerning difference.  The ground-based $V$ magnitudes are $-0.034 \pm 0.026$ brighter than the synthetic magnitudes (i.e., SD24-$GSPC = -0.034$). This comparison was constructed using only $GSPC$ magnitudes without flags. A similar comparison using the standard photometry of Stetson (2000) in M67 relative to the $GSPC$ data produces $\Delta$$V = -0.017 \pm 0.015$ (sd) from 542 stars.  
%{\it (For reference, similar comparison for M67, Stetson, Deltas are -.017 pm 0.015 and 0.008 for (PBS-$GSPC$).  Delta for Con-2204 in same sense, Con GSPC, -.034 pm 0.026, 0.009 pm 0.31.  ) }

% Can use plotone if no rotation needed.  Might try plottwo{}{} for some of these
\begin{figure}
\figurenum{4}
\includegraphics[angle=0,width=\linewidth]{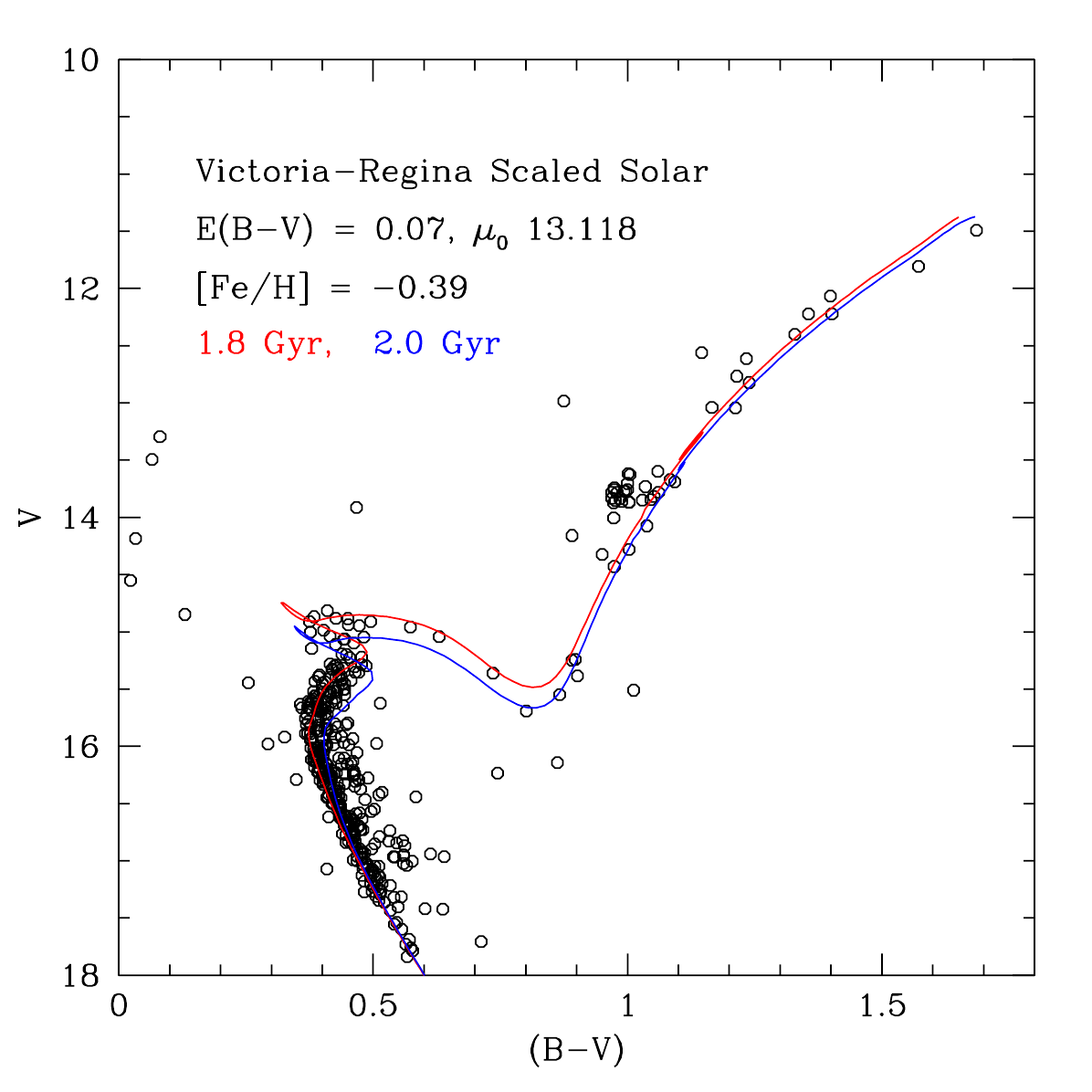}
\caption{$BV$ photometry from SD24 presented with VR isochrones, [$\alpha$/Fe]$= 0.0$, [Fe/H] $ -0.39$.  The isochrones have been adjusted for a reddenning value \ebv = 0.07 and a true distance modulus $= 13.12$, as derived from astrometry by CG20.}
\end{figure}

With few independent determinations of foreground reddening and [Fe/H], the range of values under consideration remains large, \ebv $= 0.08 \pm 0.04$ and $-0.35 \pm 0.1$. We adopt a value of \ebv = 0.07. The $BV$ photometry from SD24 for NGC 2204, filtered by membership determinations from CG18, is shown in Figure 4 with scaled-solar Victoria-Regina isochrones \citep{VR06} (VR), adopted to allow direct comparison with the results for previously analyzed metal-deficient clusters, NGC 2506 \citep{AT16, AT18b} and NGC 2243 \citep{AT21}. For consistency with the standard adopted in our previous cluster analyses using VR isochrones, we first zero the isochrones by requiring that a star of solar mass and metallicity at an age of 4.6 Gyr have $M_{V}$ = 4.84 and $B-V$ = 0.65, leading to minor adjustments, $\Delta$$V$ = 0.02 and $\Delta$$(B-V)$ = +0.013 mag. Displayed are 1.8 and 2.0 Gyr isochrones for [Fe/H]$= -0.39$ to which adjustments have been applied consistent with true distance modulus $\mu$ = 13.118  (CG20) and \ebv = 0.07. From the quality of the match at the turnoff color and the simultaneous position of the subgiant branch luminosity, the implicit age of NGC 2204 is 1.85 $\pm$ 0.05 Gyr.  Comparable fits, including the simultaneous agreement of giant branch and turnoff color, subgiant branch luminosity level, and accurate representation of the turnoff hook morphology are obtained at slightly higher reddening value (\ebv $=0.08$) for a lower metallicity ([Fe/H] $= -0.45$) and slightly lower reddening (\ebv $=0.06$) for higher metallicity ([Fe/H]$ = -0.35$). With the distance modulus largely fixed by astrometry, reddening shifts fold directly into the age determination and therefore produce morphological contradictions outside of these bounds.

\section{HYDRA Spectroscopic Analysis}

\subsection{Spectroscopic Processing With ROBOSPECT}
Since the first cluster analysis in this series for NGC 3680 \citep{AT09}, where a modest number of dwarf stars and a few giants were considered, our approach has evolved as sample sizes have grown and new tools become available. Since 2015 \citep{LB15}, we have utilized an automated line-measurement program, ROBOSPECT \citep{WH13}, to replace exclusive dependence on manual measurement of line equivalent widths (EW) as input to traditional LTE model atmosphere analysis via MOOG \citep{SN73}\footnote{Available at  http://www.as.utexas.edu/~chris/moog.html}. \citet{LB15} describe in considerable detail the various tests to which ROBOSPECT has been subjected. For the current investigation, each spectrum was individually corrected in ROBOSPECT for its unique radial velocity and run through 25 iterations of continuum fitting and line estimation using a gaussian line profile with three-$\sigma$ automatic line identification and no least-squares line deblending. All other parameters for the program were set to default values. 

The potential line-list in our chosen region (6400-6800 \AA) includes over a dozen relatively isolated iron lines and a few lines of silicon, nickel and calcium.  Lines with EW larger than 150 m\AA\ are not used, out of concern that they exceed the linear portion of the curve of growth.  Weak lines should also be excluded if their EW is near or below the anticipated EW error for a given line width and SNR, estimated from formulation originally posed by \citet{CA88} and reformulated by \citet{DP93}. For stars with typical line width and 
SNR $\sim 100$, one-sigma error is $\sim5$ m\AA, suggesting exclusion of EW values below 15 m\AA. Our final line list contains 20 lines of interest (15 Fe, 3 Ni, 1 Ca, 1 Si), presented along with the relevant atomic parameters in Table 2. All of these lines are generally present and measurable in the cooler stars but very few stars near the turnoff have as many as four iron lines above an anticipated two-$\sigma$ threshhold of 10 m\AA, therefore metal abundance determinations will not be attempted for individual stars in this class.

\floattable
\begin{deluxetable}{rcrccrr}
\tablenum{2}
\tablecaption{Lines Used with Results from Giant Stars}
\tabletypesize\small
\tablewidth{0pt}
\tablehead{
\colhead{$\lambda$(\AA)} & \colhead{Element} & \colhead{log $gf$} &
\colhead{Mdn EW} & \colhead{$N_{Star}$} &\colhead{[X/H]} & \colhead{MAD} } 
\startdata
6597.56  & Fe & -1.018 &  57 & 47 & -0.27 & 0.08 \\
6627.54  & Fe & -1.561 &  35 & 47 & -0.44 & 0.04 \\
6646.93  & Fe & -4.032 &  31 & 43 & -0.34 & 0.04 \\
6653.91  & Fe & -2.519 &  22 & 23 & -0.32 & 0.04 \\
6677.99  & Fe & -1.574 & 146 &  4 & -0.46 & 0.04 \\
6703.57  & Fe & -3.058 &  74 & 47 & -0.39 & 0.06 \\
6710.32  & Fe & -4.767 &  58 & 47 & -0.47 & 0.06 \\
6725.36  & Fe & -2.276 &  27 & 44 & -0.41 & 0.07 \\
6726.67  & Fe & -1.105 &  56 & 47 & -0.42 & 0.07 \\
6733.15  & Fe & -1.539 &  32 & 46 & -0.42 & 0.07 \\
6750.15  & Fe & -2.775 & 104 & 47 & -0.54 & 0.06 \\
6806.86  & Fe & -3.200 &  63 & 47 & -0.45 & 0.05 \\
6810.27  & Fe & -1.085 &  60 & 47 & -0.37 & 0.07 \\
6820.37  & Fe & -1.184 &  54 & 47 & -0.32 & 0.10 \\
6837.01  & Fe & -1.848 &  29 & 44 & -0.24 & 0.06 \\
6717.68  & Ca & -0.208 & 134 & 12 & -0.11 & 0.05 \\
6721.85  & Si & -1.013 &  42 & 47 & -0.26 & 0.08 \\ 
6643.63  & Ni & -1.926 & 131 & 32 & -0.54 & 0.06 \\
6767.77  & Ni & -2.159 & 114 & 42 & -0.57 & 0.06 \\ 
6772.31  & Ni & -0.948 &  66 & 47 & -0.45 & 0.07 \\
\enddata
\end{deluxetable}

\floattable
\begin{deluxetable}{rrrrrrr}
\tablenum{3}
\tablecaption{Revised H15 Color-Temperature Relations}
\tablewidth{0pt}
\tabletypesize\small
\tablehead{
\colhead{Class} & \colhead{Num} &\colhead{$(B-V)_{min}$} &\colhead{$(B-V)_{max}$} 
&\colhead{Min.$T_{\mathrm{eff}}$} &\colhead{Max.$T_{\mathrm{eff}}$} &\colhead{std.dev} \\
\colhead{ } & \colhead{$a0$} &\colhead{$a1$} &\colhead{$a2$} &\colhead{$a3$} & \colhead{$a4$} &\colhead{$a5$}  } 
\startdata
MS/MSTO &    99  & 0.21 & 1.51 & 7734 & 3618 &  139 \\
\nodata & 8099.9436 & -4343.1089 &  959.1094 &  256.6719  &  -39.6315 &     52.6089 \\
SGB/RGB & 72 & 0.43 & 1.62 & 6478 & 3602 &  100 \\
\nodata & 7702.3968 & -3601.1812 &  695.6553 &  431.7820 &  -12.5042 &  -194.4370 \\
\enddata
\tablecomments{\teff $ =  a0 + a1\cdot X + a2\cdot X^2 + a3\cdot X\cdot$[Fe/H]$ +a4\cdot$[Fe/H]$ + a5\cdot$[Fe/H]$^2$}
\end{deluxetable}

%\input{Aux/Tab4_RGFe.tex}
% print connumbers just long enough to verify -- nocolhead, hrrr
\floattable
\begin{deluxetable}{hrrrrrrr}
\tablenum{4}
\tablecaption{[Fe/H] for Giants, in order of increasing $V$ magnitude}
\tablewidth{0pt}
\tablehead{
\nocolhead{ConID} & \colhead{WOCS ID} & \colhead{[Fe/H]} & \colhead{N} & \colhead{MAD} & 
\colhead{$T_{\mathrm{eff}}$} & \colhead{\logg} & \colhead{\vturb} } 
\startdata
 C2281 & 1006 &-0.42 &14 &0.19 &3805 &1.0 &1.80 \\
 C0897 & 1017 &-0.30 &14 &0.04 &4074 &1.4 &1.72 \\
 C0589 & 1028 &-0.30 &14 &0.04 &4147 &1.5 &1.70 \\
 C3467 & 4014 &-0.41 &14 &0.08 &4069 &1.4 &1.72 \\
 C4231 & 2028 &-0.38 &14 &0.08 &4194 &1.6 &1.68 \\
 C1792 & 1011 &-0.38 &14 &0.06 &4542 &1.8 &1.64 \\
 C2309 & 1002 &-0.35 &14 &0.05 &4369 &1.8 &1.64 \\
 C3125 & 3011 &-0.40 &14 &0.05 &4405 &1.8 &1.64 \\
 C2716 & 6014 &-0.32 &14 &0.11 &4359 &1.8 &1.64 \\
 C3004 & 2009 &-0.38 &13 &0.03 &5142 &2.3 &1.72 \\
 C2405 & 3006 &-0.45 &13 &0.07 &4502 &2.0 &1.60 \\
 C2806 & 2007 &-0.36 &13 &0.03 &4411 &2.0 &1.60 \\
 C3599 & 2016 &-0.50 &13 &0.08 &4721 &2.4 &1.62 \\
 C2064 & 1003 &-0.33 &14 &0.07 &4850 &2.4 &1.62 \\
 C1479 & 1010 &-0.41 &14 &0.07 &4841 &2.4 &1.62 \\
 C2216 & 1004 &-0.41 &14 &0.05 &4670 &2.4 &1.62 \\
 C2501 & 2003 &-0.40 &14 &0.08 &4651 &2.4 &1.62 \\
 C2939 & 2010 &-0.35 &13 &0.09 &4852 &2.4 &1.63 \\
 C2028 & 7014 &-0.60 &11 &0.10 &4775 &2.4 &1.62 \\
 C4001 & 2023 &-0.34 &13 &0.07 &4911 &2.5 &1.60 \\
 C1620 & 1019 &-0.47 &13 &0.08 &4852 &2.5 &1.59 \\
 C2153 & 3003 &-0.40 &13 &0.06 &4911 &2.5 &1.60 \\
 C2549 & 3009 &-0.44 &13 &0.06 &4868 &2.5 &1.59 \\
 C4156 & 8027 &-0.37 &13 &0.04 &4859 &2.5 &1.59 \\
 C2847 & 8014 &-0.48 &13 &0.04 &4922 &2.5 &1.60 \\
 C4165 & 4733 &-0.41 &14 &0.06 &4719 &2.4 &1.62 \\
 C4426 & 1736 &-0.51 &13 &0.06 &4897 &2.5 &1.59 \\
 C2397 & 3005 &-0.36 &13 &0.04 &4922 &2.5 &1.60 \\
 C1113 & 3015 &-0.34 &14 &0.07 &4886 &2.5 &1.59 \\
 C2077 & 4003 &-0.42 &13 &0.08 &4752 &2.5 &1.58 \\
 C2356 & 4007 &-0.68 &10 &0.15 &4788 &2.5 &1.58 \\
 C1747 & 5008 &-0.47 &13 &0.04 &4906 &2.5 &1.59 \\
 C4465 & 7028 &-0.33 &14 &0.06 &4877 &2.5 &1.59 \\
 C1998 & 4018 &-0.35 &14 &0.08 &4846 &2.5 &1.59 \\
 C3182 & 3016 &-0.39 &14 &0.06 &4846 &2.5 &1.59 \\
 C2967 & 6008 &-0.36 &13 &0.06 &4913 &2.6 &1.56 \\
 C0257 & 8028 &-0.36 &13 &0.06 &4913 &2.6 &1.56 \\
 C2105 & 5003 &-0.30 &14 &0.04 &4769 &2.6 &1.54 \\
 C0637 & 6021 &-0.34 &14 &0.06 &4846 &2.7 &1.51 \\
 C1092 & 5015 &-0.54 &12 &0.02 &4966 &2.8 &1.49 \\
 C2425 & 8011 &-0.42 &13 &0.06 &4911 &2.8 &1.48 \\
 C2984 & 8008 &-0.37 &11 &0.04 &5090 &3.3 &1.32 \\
 C0860 & 9020 &-0.36 &14 &0.09 &5104 &3.2 &1.37 \\
 C2507 & 8005 &-0.40 &13 &0.10 &5078 &3.2 &1.36 \\
 C1584 & 14011 &-0.27 &14 &0.10 &5161 &3.3 &1.34 \\
\enddata
\end{deluxetable}

To minimize external variables that might result from fiber-to-fiber variations and/or secular changes in the instrument's sensitivity, 
we reevaluate solar \loggf\ values for each run and spectrograph setup by tuning those values to recreate solar abundances in daytime sky spectra, obtained in the appropriate fiber configuration for that purpose.  Our evaluation of EW values to produce line-by-line abundance estimates is carried out in the context of model atmospheres generated by linear interpolation between Kurucz \citep{KU95} atmospheric models using MOOG's \citep{SN73} $abfind$ driver.  We employ the 2014 LTE version of MOOG for which the solar iron abundance is set at 7.50.  In subsequent discussions, the solar A(Fe)\footnote{A(X)=log$N_{X}$ - log$N_{H}$ + 12.00}
value will be subtracted from the MOOG abundance for each determination yielding [Fe/H] relative to the sun.  For solar abundance checks, we employ a model constructed with (\teff, \logg, \vturb) = 5770 K, 4.40, 1.14 \kms, where \vturb\ denotes the microturbulent velocity parameter.

\subsection{Atmospheric Parameters: \teff, Surface Gravity, and Microturbulent Velocity}
In the past we have used color-temperature relations from \citet{RA05} and \citet{DE02} for giants and dwarfs respectively. Because of its rich sample of over 150 stars and a homogeneous compilation of direct temperatures, \citet{H15} (H15) will be adopted going forward, with a few modifications.  We want to prioritize disk metallicities and stars inside a \teff\ range of 3600 to 7800 K, so some stars outside those limits were eliminated. We also reassigned luminosity class for the stars using absolute magnitudes with the help of $Gaia$ parallaxes and photometric values from H15. The relations from H15 do not cover likely subgiants. 

159 stars were sorted into clear main sequence stars (MS), evolved stars near the MSTO, probable subgiant (SG) stars and clear RGB stars.  To provide adequate color range and continuity between these classes, fits between \teff, unreddened $(B-V)$ color and [Fe/H] were made for 99 stars comprising 87 MS stars and 12 MSTO stars, then for 72 stars including 60  RG, 6 SG and the evolved stars near the MS. The fits, with dependent variables of color and metallicity, are patterned after those used by H15 except that \teff\ rather than $\theta = 5040/$\teff\ is solved for. The equations for RG/SG and MS/MSTO classes, color and temperature limits are summarized in Table 3. 

For our Hydra analysis, photometry from SD24 is used for all except a few stars outside the areal coverage of that study; for these the synthesized $GSPC$ photometry has been adopted.  A reddening of \ebv $ = 0.07$ and [Fe/H] = $-0.4$ were chosen for the color-temperature conversions.   

Surface gravity estimates (\logg) were obtained by direct comparison of $V$ magnitudes and $(B-V)$ colors to the scaled-solar VR isochrones with [Fe/H]$=-0.39$ near 1.9 Gyr in age. For stars not immediately adjacent to the isochrone curve, adjustments were made to the nearest isochrone point of similar color-temperature based on the magnitude difference between the isochrone and the star such that $\Delta$log $g = \Delta$ log $V/ 2.5$. For the comparison, $(m - M)_{0}$ = 13.118 and \ebv = 0.07 were applied to a 2.0 Gyr isochrone. Input estimates for \vturb\ were constructed using the formulation by \citet{Br12} for stars within the valid ranges of \logg\ and \teff.  For the most luminous giants with \logg\ $\leq 2$,  \vturb = $2.0 -0.2$ \logg.

Comparisons of our \teff\ determinations relative to past work have been constructed, beginning with the infrared photometric studies by \citet{HFC}.  These authors explored $JHK$ colors to establish clearer pictures of clusters' reddening, distances, ages and metallicities, developing profiles of cluster giant branches and building on pre-established indications of [Fe/H] and reddening.  In this context, they studied giants in NGC 2204, making use of the earlier estimates of 0.08 for \ebv and $-0.35$ for [Fe/H] from photometric studies by \citet{D81} and \citet{H76a}.  As such, this is less an independent confirmation of our color-temperature scale than a check on it based on longer wavelength photometry.  From seven giants that are both cluster members and within the temperature bounds of the revised H15 color-temperature relation, the \citet{HFC} temperatures are $30 \pm 71$ K (sd) warmer.

Similarly, \citet{JA11} provide atmospheric parameters for a group of 13 giants for which the median [Fe/H] is $-0.26$.  The temperature scale is not independently determined from the spectra, but is based on photometry, a literature value for \ebv $=0.08$  and the color-temperature relation of \citet{al99}.  The \teff\ values for \citet{JA11} are on average $5 \pm 93$ K cooler than those derived herein from the revised H15 color-temperature calibration and $BV$ photometry.

The spectroscopic study by \citet{CA16} is significant because the authors were able to derive \teff\ entirely from spectroscopy in the established manner {\it i.e.}, normalizing the trend of A(Fe) with excitation potential.  They also had sufficient spectral coverage and resolution to include both neutral and ionized iron lines, enabling an independent derivation of \logg.  

%Excluding the likely binary, W4119 = WOCS1003,  our \logg\  values are lower than those of \citet{CA16} by a modest 0.26 on average. An offset of this size has a minor impact on our abundances and no effect on our conclusions. By contrast, comparison of \citet{CA16}'s derived temperatures with $(B-V)$ colors shows a very similar slope to the H15 color-temperature relation but considerably hotter, with two outliers. {\bf This result is not dependent on the source of the (B-V) photometry, \citet(K97) or SD24,  and suggests a potential discrepency between \teff\ and photometric color for these two stars.} If those outliers (WOCS4007 and 7014) are excluded, the average difference in \teff\ in the sense (RevisedH15-Carlberg) $=-152 \pm 33$ K. We note that if the H15 color-temperature relation were to be preserved at [Fe/H]$=-0.40$, the foreground reddening \ebv\ would have to be considerably higher, nearly 0.14, to produce temperatures this high.

Excluding the likely binary, W4119 = WOCS1003,  our \logg\  values are lower than those of \citet{CA16} by a modest 0.26 on average. An offset of this size has a minor impact on our abundances and no effect on our conclusions. 
By contrast, comparison of \citet{CA16}'s derived temperatures for most stars with $(B-V)$ colors, whether from \citet{K97} or SD24, shows a very similar slope to the H15 color-temperature relation but considerably hotter.  If the worst outliers (WOCS4007 and 7014) are excluded, the average difference in \teff\ in the sense (RevisedH15-Carlberg) is $-152 \pm 33$ K. We note that if the H15 color-temperature relation were to be preserved at [Fe/H]$=-0.40$, the foreground reddening \ebv\ would have to be considerably higher, nearly 0.14, to produce temperatures this high.  The two outliers for which a potential discrepancy between \teff\ and photometric color is suggested will be revisited later in this paper. 

By contrast, comparison of atmospheric parameters for some of the brighter cluster stars as determined in the $Gaia$ DR3 parameter pipeline GSP-phot \citep{andr} are much harder to understand.  The GSP-phot \teff\ values have a less coherent relationship with $(B-V$) colors and are, on average, $\sim 200 \pm 266$ K higher, although the large dispersion in the individual differences makes this statistic hard to employ.  $Gaia$ GSP-phot also derives \logg, nearly identical to values derived here from isochrones and photometry but with a sizeable standard deviation of 0.36 on that nearly null difference.

\subsection{Metal Abundance Determinations }
Model atmospheres were constructed for member giants using the grid of ATLAS9 models crafted by \citet{KU95} for the input \teff, \logg, \vturb\ listed in Table 4. We continue to use Kurucz models, though we have taken advantage of expanded model availability with [Fe/H] values now extending from $-5.0$ to $1.0$.  Our access to the models was via the on-line repository\footnote{http://kurucz.harvard.edu/grids/ } 
 which was updated in 2014 and 2018.  We construct models from simple linear interpolation of models with bracketing \teff\ and \logg\ values.  
 
Each star's measured equivalent widths and model serve as input to the {\it abfind} routine of MOOG to 
produce individual [A/H] estimates for each measured line for each star, producing hundreds of separate abundance measures for single member stars of NGC 2204 with up to 15 [Fe/H] values for each star. 

Our discussion utilizes median statistics to avoid the awkwardness of averaging logarithmic quantities. The MAD, or median absolute deviation, is a robust indicator of dispersion among individual values. For typical distributions without extreme outliers, a traditional standard deviation $\sim 1.48 \times$ MAD. 

\begin{figure}
\figurenum{5}
\includegraphics[angle=0,width=\linewidth]{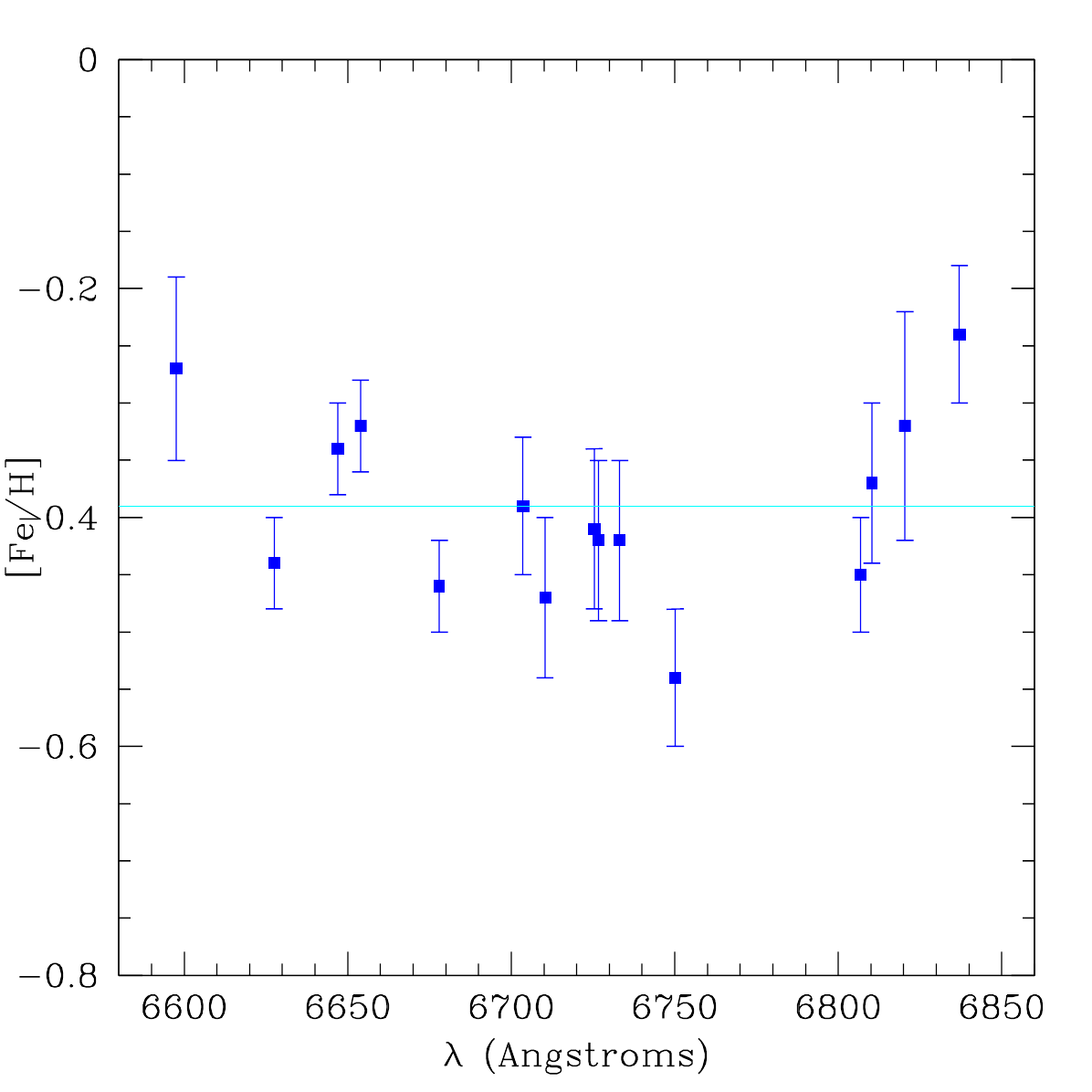}
\caption{Abundance estimates for individual Fe lines.  Results are shown for RG/SG members only.  Errorbars show MAD statistics} 
\end{figure}
\begin{figure}
\figurenum{6}
\includegraphics[angle=0,width=\columnwidth]{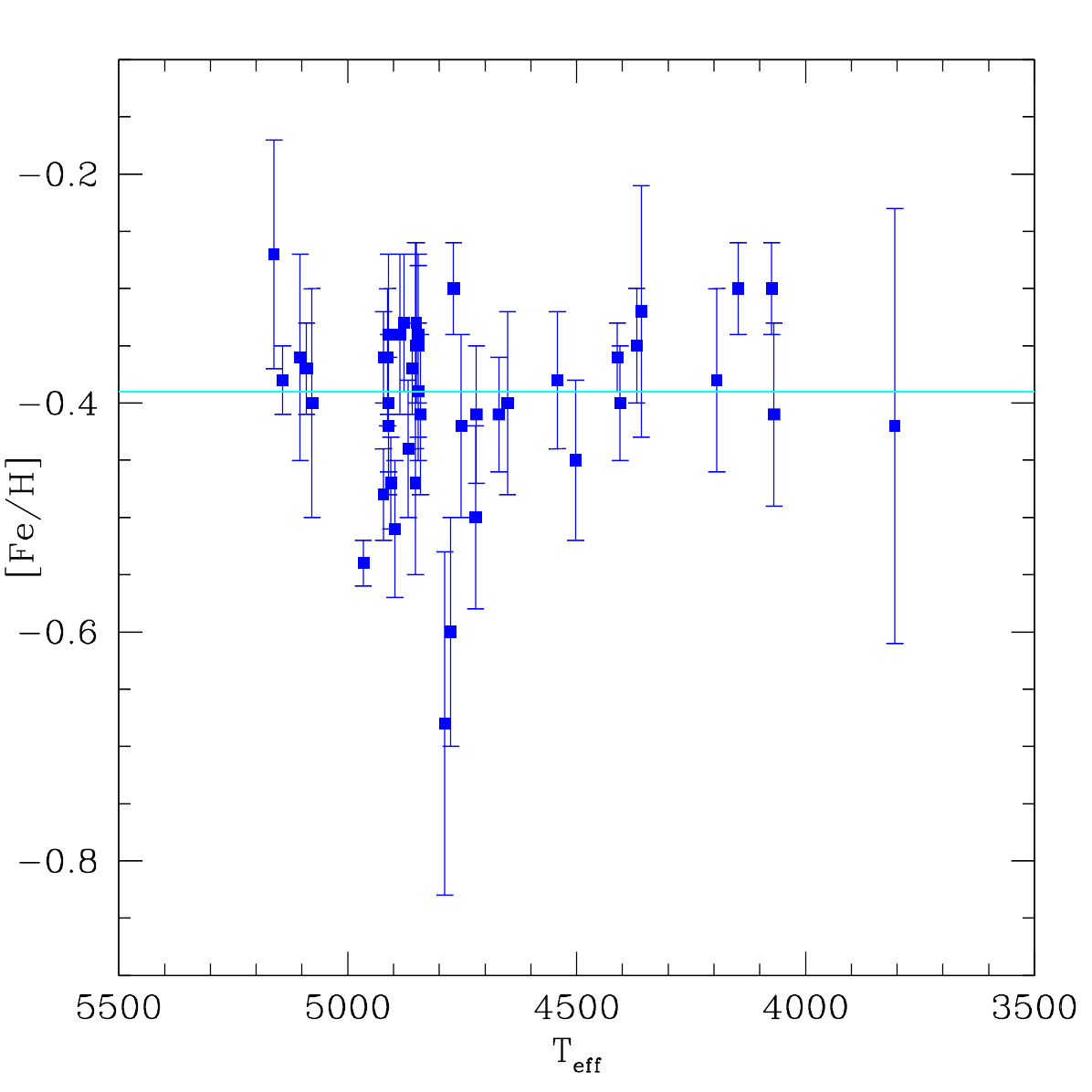}
\caption{Estimates for [Fe/H] for individual RG/SG stars as a function of \teff.  Errorbars show the MAD statistic.}
\end{figure}

Our large number of 703 [Fe/H] estimates from 45 evolved stars is reduced to 627 if EW values below 15 or above 150 m\AA\ are excluded from further analysis.  As is typical for median statistics considering large samples, modest changes in filtering have little consequence.  The median [Fe/H]  based on all eligible lines is $-0.40 \pm 0.08$, where the error quoted is the MAD statistic, implying a more traditional standard deviation of 0.12.  If the median of all 45 separate stellar abundance estimates is considered, the results are predictably similar: $-0.38 \pm 0.06$ (MAD) for EW measures meeting the cutoffs described above.
Presentation of results in Table 4 includes the \teff, \logg, and \vturb\ estimates for each evolved member star, while results specific to individual lines were included in Table 2.
Among the evolved stars summarized in Table 4, two stars happened to be observed in separate configurations, spanning both years, 2014 and 2015, providing an opportunity to verify the repeatability of our [Fe/H] determinations.  Results for these two stars differ by $\sim 0.05$ dex.

In Figures 5 and 6, we illustrate the Fe abundance information by line wavelength and stellar \teff.  For the 45 giants discussed above, stellar abundance estimates are based on 10 to 14 lines, with 13 lines being the median value.  Two of the lowest [Fe/H] values occur for stars with fewer than 12 lines measurable.  These stars, WOCS7014 and WOCS4007, have atypically large MAD values and are also the two stars previously noted to have spectroscopic temperatures inconsistent with their $(B-V)$ colors, according to \citet{CA16}. If, as indicated by those spectroscopic temperatures, photometry assigns temperatures to these stars that are too low, a spuriously low metal abundance would be derived as a result, by as much as 0.15 dex for the star with the most discrepant temperature.  We reiterate that this discrepancy between spectroscopic and photometric temperatures is not the exclusive result of use of photometry from \citet{con}; \citet{CA16} uses photometry from \citet{K97} and notes the same issue.  We further stress that retention of results for these two stars does not affect the overall abundance results for the cluster, due to the use of median statistics.

Our limited results for elements other than Fe are included in Table 2, with values for [m/H] accompanied by the number of evolved stars' lines and the MAD statistic for each median value.  The relatively small number of included star lines for Ca is due to the typically large EW for this line in cooler stars.  The results for Ca and Si imply slight $\alpha$-enhancement.  For a cluster [Fe/H]$=-0.40$, [Si/Fe]$= 0.14 \pm 0.12$ and [Ca/Fe]$=0.29 \pm 0.07$, where the errors noted here are standard deviations based on the MAD statistics.  The median value for [Ni/H] from all lines is $-0.52$, implying [Ni/Fe] = $-0.12 \pm 0.10$ (sd). With similar precepts for the determination of atmospheric parameters, ROBOSPECT was run on the more than 50 MS and MSTO stars. Unfortunately, the warm, relatively metal-weak and often broadened spectra produced far fewer than 15 succesfully measured Fe lines in all but 14 stars.  Even for these, only 4 to 6 lines were measurable, suggesting that a star-by-star analysis would be ineffective. Considering all of the Fe determinations {\it en masse}, from 50 eligible Fe lines among the 14 stars, a median value of $-0.19$ was derived for [Fe/H] with a large MAD statistic of 0.39.  At best, these results may be considered weakly consistent with the larger sample and clearer results from SG and RGB stars.

What are the consequences of an incorrectly adopted reddening on the derived [Fe/H]? An inappropriately high reddening value assigns a higher than needed \teff\ to the star, forcing a higher derived [Fe/H] value.  For giant stars with [Fe/H] values near $-0.40$, the sensitivity of derived [Fe/H] values to adopted \ebv\ is $\sim 0.015$ for each increment of 0.01 in adopted \ebv. Fortunately, derived [Fe/H] values are less sensitive to the accuracy of the model atmospheres' assumed [Fe/H] values: an increment of 0.2 in model [Fe/H] produces a spurious increment of $\leq 0.05$ in derived [Fe/H].

\section{Li: Abundance Estimation and Evolutionary Implications}
\subsection{Li From Spectrum Synthesis}

The challenge in estimating the abundance of Li is dominated by the fact that only one line near 6707.8 \AA\  is accessible for analysis. While this isn't a problem for lines of adequate strength, for hotter stars where the line strength weakens with increasing \teff\  and/or for stars with rapid rotation where line broadening creates a wide but shallow profile difficult to define relative to the continuum, measuring the equivalent width can be an exercise in futility that, at best, only permits upper limits. Adding to the challenge is the existence of a weak Fe I line at 6707.4 \AA\  which, with inadequate resolution, can blend with the Li line creating an enhanced equivalent width for Li if not accounted for. Fortunately, this last issue is of minor importance due to the low metallicity of NGC 2204.

While the Li line's equivalent width can be measured non-interactively using ROBOSPECT or interactively using $splot$ in IRAF, as in past analyses in the cluster series, we have chosen to use spectrum synthesis to define the individual stellar abundances, a critical approach for stars where the Li line is weak to nonexistent. The procedure as laid out in previous papers is as follows: each candidate star's spectrum is compared to the relevant model atmosphere using the $synth$ driver in MOOG. When the model \teff\ has been appropriately chosen, lines other than Li show consistent levels of agreement between spectrum and model, where ``consistency'' presumes that the line profile characterizing the spectrum is also appropriately modeled in MOOG. The line profile incorporates the known instrumental line width (0.55 m\AA\ for Hydra spectra) as well as corrections for limb-darkening (coefficient taken uniformly to be 0.5) and broadening due to rotation, for which projected rotational velocities may be interactively altered. 

For all completed syntheses in the Li line region, the rotational velocity which appears to best fit the observed spectrum was recorded as well as the resulting A(Li) value.  
These \vrot\ values were uniformly consistent with values obtained directly from the $fxcor$ procedure described in Section 2.2 that determines both radial and rotational velocity parameters for values between 15 and 40 \kms. For the low end of the velocity scale the instrumental resolution limits the measurable velocities to between 5 and 10 \kms.  Projected rotational velocities determined from $fxcor$ analysis for all stars with Li detections or estimated upper limits are included with the A(Li) values presented in Table 5.

For cases in which the Li line is distinct and clearly present, an abundance for Li is discerned by comparing the fit for different A(Li) values. In many cases, the Li line is not obviously distinguishable from noise or nearby spectral features; in these cases, an upper limit is determined by noting the value of A(Li) below which changes to the spectrum-model fit are no longer distinguishable.  A(Li) determinations, including upper limits, for all measurable spectra are listed in Table 5. 

\subsection{Li Evolution and the CMD: Evolved Stars}

Figure 7 shows the CMD for evolved stars with Li detections and Li upper limits. Two stars, WOCS1005 and WOCS2006 (open circles), require some clarification since synthesis was not possible due to the stars' extremely cool temperatures.  WOCS2006, flagged for potential binarity and variability, is the reddest star in the sample, past the ``dog-leg'' of the giant branch.  Although too cool for synthesis, it does have a conspicuous Li line with an EW of over 240 m\AA.  WOCS1005 has a similar spectrum with dominant molecular bands, yet no obvious Li line. 

The next coolest star, WOCS1017, has a measurable though more modest Li EW of 29 m\AA, but synthesis appears to suggest an implausibly low upper limit.  The other cool stars for which Li was not detected all have Li line EWs from ROBOSPECT of 10 m\AA\ or less if the contribution of the adjacent Fe line is estimated and subtracted. 

\begin{figure}
\figurenum{7}

\includegraphics[angle=0,width=\linewidth]{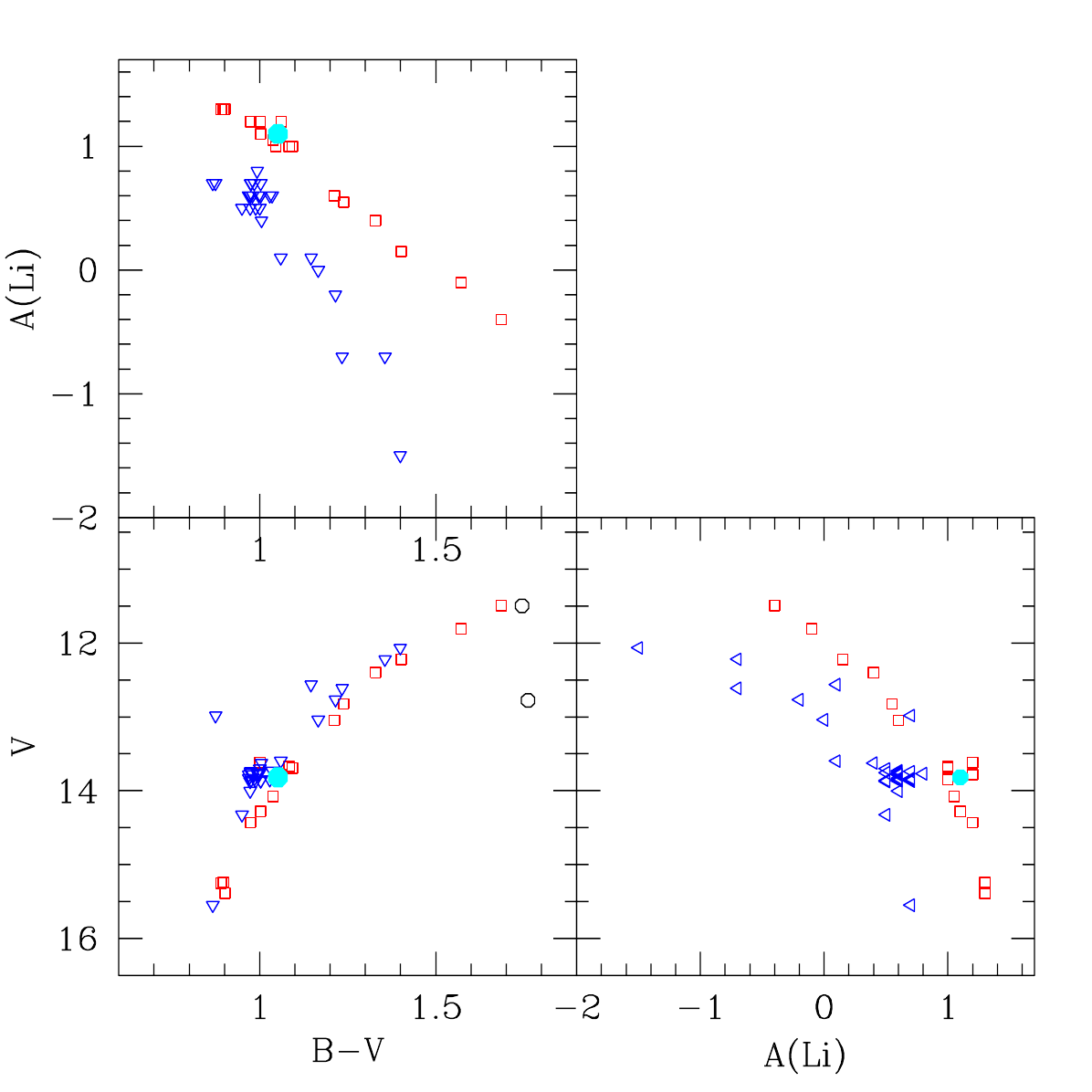} 
\caption{Li abundance patterns among the giants. The location of WOCS4009 is shown with a cyan dot. Blue triangles indicate upper limits to A(Li) while red squares illustrate detections. }
\end{figure}

One of the striking features of Figure 7 is the separation of the red giants into clear detections (red squares) versus stars for which only upper limits are derivable (blue triangles). All the stars that define the red boundary in the CMD delineating the first-ascent red giant (FRG) branch have measurable Li, starting with A(Li) just below the canonical prediction of 1.4 at the base of the giant branch and declining to -0.4 near the tip of the giant branch. While it may go lower beyond this point, as discussed earlier, the two coolest stars fall outside the \teff\  limits of the synthetic spectra. By contrast, with only one exception, the red clump giants (RCG) only have upper limits centered on A(Li) = 0.6 $\pm$ 0.2 and typically 0.5 dex below the measured detections for FRG at the same luminosity. Thanks to the precision photometry of SD24, for stars brighter than the RCG the Li bifurcation translates into a color separation where the bluer stars all have distinctly lower A(Li) limits than the detections on the RCG at the same luminosity. 

It should be emphasized that this bifurcation is not an artifact of the spectrum synthesis approach. The separation for the giants above the clump is at least as distinct if only Li EW are considered.  For the stars with Li upper limits, EW are uniformly low, 20 to 26 m\AA. This measured EW includes the contribution from the nearby Fe line, with an estimated strength of 10 to 20 m\AA\ going up this part of the RGB.  In contrast, the six giants with Li detections have Li EW from 60 to 120 m\AA, larger for the cooler stars. The separation in the CMD of these giant classes above the level of the RGB clump is subtle but believable, leading to the prediction that the bluer stars are post-RGC, asymptotic giant branch stars (AGB) evolving back up the giant branch. In support of this interpretation, we note that one of the upper limit stars, WOCS3011 = W2212, was selected for observation by \citet{CA16} with the expectation that it was an AGB star, an assignment confirmed by their determination of the AGB separation in the \teff, \logg\ diagram. Unfortunately, no other stars among our group of cool stars with upper limits for Li were observed by \citet{CA16}.

\floattable
\begin{deluxetable}{hrrrrhrrrr}
\tablenum{5}
\tablecaption{Synthesis Results for Lithium}
%\tabletypesize{small}
\tablewidth{0pt}
\tablehead{
\nocolhead{ConID} & \colhead{WOCS ID} & \colhead{\vrot} & \colhead{$\sigma_{vrot}$}  & \colhead{A(Li)} &
\nocolhead{ConID} & \colhead{WOCS ID} & \colhead{\vrot} & \colhead{$\sigma_{vrot}$}  & \colhead{A(Li)} }
\startdata
2541 & 2014 & 11.3 & 0.9 & -0.40       & 3182 & 3016 &  10.7 & 0.3 & $\leq$ 0.70 \\
2281 & 1006 & 4.4 & 0.3 & -0.10        & 2967 & 6008 & 11.1 & 0.3 & $\leq$ 0.50 \\
897  & 1017 &  5.0 & 0.2 & $\leq$ -1.50   & 257 & 8028 & 13.9 & 0.4 & $\leq$  0.60 \\
589  & 1028 & 12.8 & 0.4 & $\leq$  -0.70  & 2105& 5003 & 16.1 & 0.4 &   1.05 \\
3467 & 4014 & 13.9 & 0.6 &  0.15        & 637 & 6021 & 18.0 & 0.4 &   1.10 \\
4231 & 2028 & 12.1 & 0.4 &  0.40          & 1092& 5015 & 19.8 & 0.5 & $\leq$ 0.50 \\
1792 & 1011 & 14.3 & 0.4 & $\leq$  0.10  & 2425 & 8011 & 21.1 & 0.5 &  1.20 \\
2309 & 1002 & 15.1 & 0.5 & $\leq$  -0.70  & 2984 & 8008 & 17.8 & 0.4 &  1.30 \\
3125 & 3011 & 17.5 & 0.5 & $\leq$  -0.20  & 860 & 9020 & 18.6 & 0.5 &   1.30 \\
2716 & 6014 & 12.6 & 0.4 &   0.55       & 2507 & 8005 & 19.9 & 0.5 &   1.30 \\
3004 & 2009 & 19.2 & 0.5 & $\leq$  0.70   & 1711 & 12007 & 30.2 & 1.2 &  2.80 \\
2405 & 3006 & 13.9 & 0.4 & $\leq$  0.00  & 1433 & 15013 &  41.8 & 1.7 &   2.75 \\
2806 & 2007 & 12.8 & 0.4 &   0.60      & 1096 & 12015 & 30.3 & 1.3 &   3.00 \\
3599 & 2016 & 12.9 & 0.4 & $\leq$  0.10  & 1584 & 14011 & 20.8 & 0.4 & $\leq$ 0.70 \\
2064 & 1003 & 10.4 & 0.3 &  1.20      & 3085 & 10019 & 41.1 & 1.6 &   2.85 \\
1479 & 1010 & 15.7 & 0.4 & $\leq$ 0.40  & 925 & 15024 & 20.6 & 0.6   & 2.10 \\
2216 & 1004 & 14.0 & 0.4 &  1.00     & 2241 & 13004 & 36.5 & 2.5&   2.70 \\
2501 & 2003 & 12.8 & 0.4 &  1.00     & 3992 & 20024 & 23.6 & 0.8 &   2.80 \\
2939 & 2010 & 17.4 & 0.4 & $\leq$  0.50  & 3419 &19014 &  10.4 & 0.3 &  2.70 \\
2028 & 7014 & 11.0 & 0.4 & $\leq$  0.60  & 3755 & 33024 & 47.6 & 2.9&   3.00 \\
4001 & 2023 & 12.8 & 0.4 & $\leq$  0.70  & 863 & 19023 & 25.3 & 1.1 &   2.75 \\
1620 & 1019 & 16.7 & 0.4 & $\leq$  0.50  & 1512 & 17026 &  23.8 & 1.0 &   2.80 \\
2153 & 3003 & 15.2 & 0.4 & $\leq$  0.60  & 5429 & C    & 21.2 & 1.0 &   2.80 \\
2549 & 3009 & 15.1 & 0.4 & $\leq$  0.80  & 2682 & 20012 & 21.1 & 1.0 &   2.70 \\
4156 & 8027 & 14.8 & 0.4 & $\leq$  0.60  & 3460 & 24014 & 18.6 & 0.7 &   2.85 \\
2847 & 8014 & 17.5 & 0.5 & $\leq$ 0.60  & 2927 & 19020 & 20.2 & 0.8 &   2.95 \\
4165 & 4733 & 15.3 & 0.4 &  1.20     & 1791 &  17008 & 21.6 & 1.2 &   3.00 \\
1866 & 4009 & 35.0 &  &  1.10          & 2558 & 16017 &   22.8 & 0.8 & $\leq$  2.10 \\
4426 & 1736 & 15.8 & 0.4 & $\leq$  0.60  & 1653 & 20009 & 27.6 & 1.1 &   2.60 \\
2397 & 3005 & 15.9 & 0.4 & $\leq$  0.60  & 814 & 24021 & 25.7 & 1.8 & $\leq$  2.30 \\
1113 & 3015 & 13.2 & 0.3 & $\leq$  0.70  & 2987 & 20008 & 32.7 & 1.2 & $\leq$  2.10 \\
2077 & 4003 & 19.7 & 0.6 &   1.00      & 690 & 27021 & 26.6 & 0.9 & $\leq$  2.10 \\
2356 & 4007 & 11.3 & 0.5 & $\leq$  0.60  & 3544 & 29018 & 18.8 & 0.6 & $\leq$  2.00 \\
1747 & 5008 & 13.2 & 0.4 & $\leq$  0.70  & 1639 & 25011 & 13.5 & 0.6 & $\leq$  2.40 \\
4465 & 7028 & 8.40 & 0.2 & $\leq$  0.50  & 4289 & 36028 &26.8 & 1.4 & $\leq$  2.10 \\
1998 & 4018 & 19.2 & 0.5& $\leq$  0.60  & 201 &  E  &  21.0 & 0.8 & $\leq$  2.00 \\
\enddata
\end{deluxetable}

The results presented in Figure 7 include photometry and a Li abundance for WOCS4009 (cyan filled circle), for which the only spectroscopic data are those of the VLT spectrum shown in Figure 3. Because of its location in the CMD between the RGC and the FRG it is impossible to determine which category describes the star. Although synthesis is difficult for such wide lines, an abundance of A(Li) = 1.1 is suggested by comparison to a model spectrum broadened by 35 \kms. The model was constructed in the same manner as the other giants adopting \teff$= 4743$ K, \logg$= 2.5$ and \vturb $= 1.58$ \kms.  Plausible estimates for the actual equatorial rotational velocity for WOCS4009 are considerably higher.  For a radius of $\sim 12$ $R_{\sun}$, as suggested by the isochrones, and the observed period from ASAS-SN of 6.18 days, an equatorial rotational velocity $\geq 100$ \kms\ is implied.  If WOCS4009 is a member of the RGC, it is clearly Li-rich compared to the typical RGC star, though its record as a photometric variable and potential radial velocity variable  make its history and evolutionary status tricky to discern. It appears to be another example of a growing class of giant stars with discordant Li abundances found in anomalous positions in the CMD and/or with broad-lined spectra and photometric variability associated with rapid rotation, such as W7017 in NGC 6819 \citep{AT13}, star 4128 in NGC 2506 \citep{AT18b}, star W2135 in NGC 2243 \citep{AT20, AT21}, and, most recently, star 4705 in NGC 188 \citep{SU22}. To date, only W7017 has been confirmed as a low mass RGC star, despite its position in the CMD redward of the FRG branch \citep{CA15}. With a luminosity at the level of the RGC, a broad-line spectrum, and photometric variability, supposedly due to rotation, on timescales of 6-7 days, WOCS4009 closely resembles star W2135 in NGC 2243.

\subsection{Li Evolution and the CMD: MSTO Stars}

Figure 8 shows the trends of A(Li) with $(B-V)$ color and $V$ magnitude for MSTO stars in NGC 2204.  The stars for which synthesis could not be carried out are designated with open black symbols, while the other symbols have the same meaning as in Figure 7. For the stars with usable spectra, the A(Li) pattern is straightforward with A(Li) remaining constant at 2.85 $\pm$ 0.15 between $V$ = 15.5 and 16.5. The only exception is the one star evolving into the subgiant branch with A(Li) $\sim$ 2.1 {\it en route} to the base of the giant branch 
by which time its A(Li) will have dropped to just below 1.4.
However, at the critical boundary of $V$ = 16.5, one sees the distinct transition to the Li-dip, bottoming out at an upper limit of A(Li) = 2.0. 

In the similar anticenter cluster, NGC 2243, the center of the main sequence Li-dip is located at $\sim 1.15 \pm 0.02$ \msun, with the transition to undepleted Li abundances on the high mass end, the ``wall'', appearing over the range between 1.21 and 1.24 \msun.  Extrapolating these masses for a slightly more metal-rich cluster ($\Delta$[Fe/H] = +0.15 dex) implies adding 0.06 \msun\ to these feature masses based upon an [Fe/H]-based slope of 0.4 \citep{AT21}. The expectation is that the wall in NGC 2204 should be positioned between 1.27 and 1.30 \msun.  Using the VR isochrones shown in Figure 4, the mass associated with each $V$ magnitude at the MSTO is plotted along the central vertical axis of Figure 8. While the number of stars defining the edge is limited in comparison with the comparable plot for NGC 2243, it is apparent that the transition between $V$ = 16.5 and 16.6 is consistent with the predicted mass range and clearly higher than the mass identified for the more metal-deficient NGC 2243. The expected low-mass limit of the Li-dip should occur near 1.15 \msun,  fainter than the magnitude limit of the current survey, as confirmed by Figure 8.

\begin{figure}
\figurenum{8}
\includegraphics[angle=0,width=\linewidth]{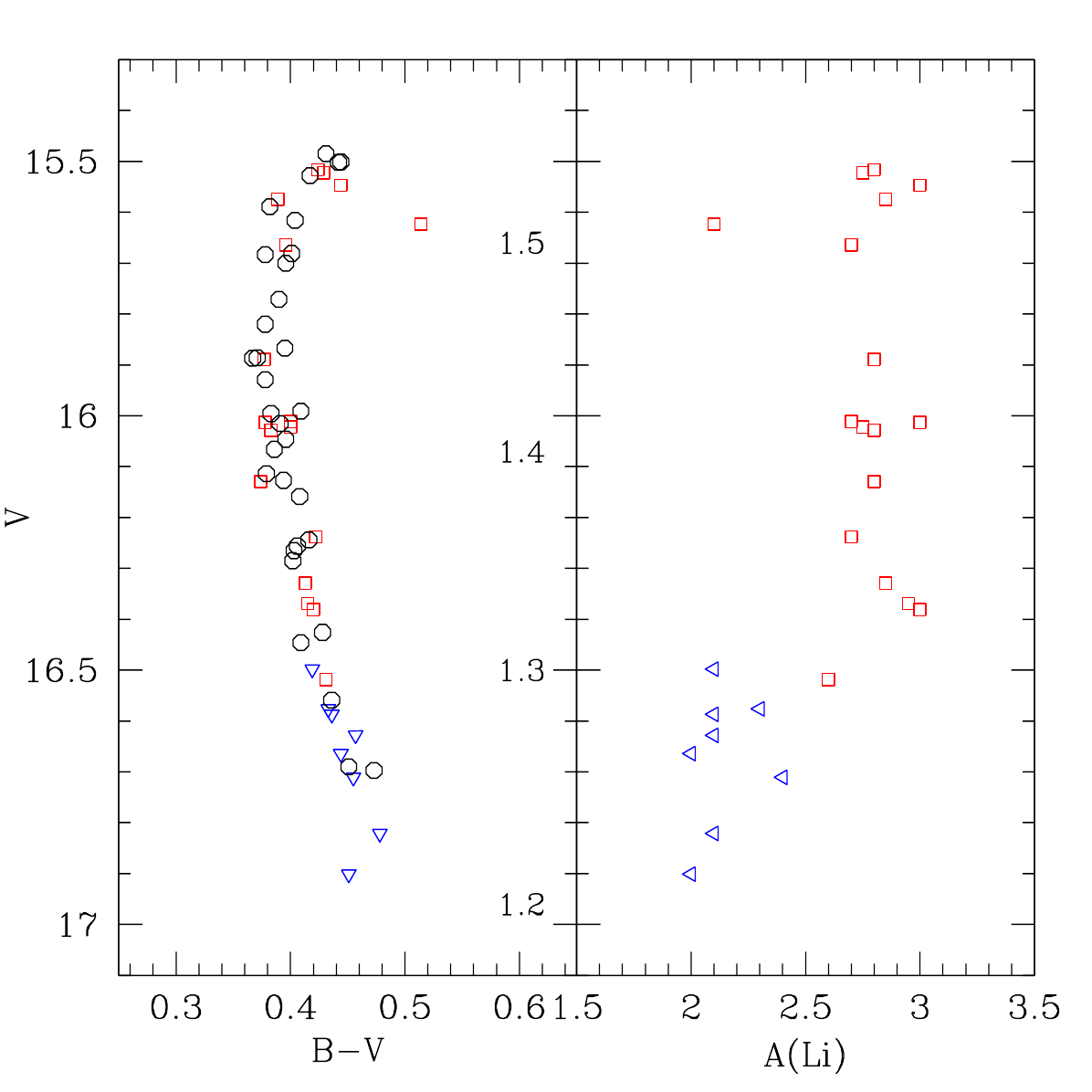}
\caption{Li abundance patterns among MSTO stars.  As before, red and blue  symbols show stars with detected Li or upper-limits.  Black symbols show the CMD location for stars with no successful synthesis measurement of Li.  Numbers to the right of the CMD in the left panel show the stellar masses in solar units at the corresponding magnitude levels, based on isochrones of Figure 4.}
%Arrows show the luminosity levels associated with the projected location of the Li-dip center and the limits over which the ``wall'' feature is predicted to appear.}
\end{figure}

\subsection{Li Evolution: NGC 2204 and the Li Cluster Pattern} 
The Li trends with mass and evolutionary state, as laid out in Figures 7 and 8, are in qualitative agreement with expectations as delineated in past analyses of clusters of similar age and/or metallicity. Stars above the Li wall at the MSTO have a modest range in abundance but generally follow a distribution maxed out between A(Li) = 2.9 and 3.4. Stars leaving the main sequence to the subgiant branch exhibit a steep decline in A(Li) with color, approaching a value typically below A(Li) = 1.4 at the base of the giant branch. Between this point and the luminosity level of the RGC, the FRG branch shows at most a minor decline of $\sim$0.1 dex in A(Li). Beyond this point, A(Li) drops at an increasing rate as one approaches the tip of the FRG branch. Beyond this evolutionary point, the RGC stars, with a few key exceptions, exhibit only upper limits to A(Li), limits which fall below the measured values for the FRG at the same luminosity, in theory due to the prior effects of the He-flash at the FRG branch tip.

The role of metallicity in defining the mass range of the Li-dip has already been noted. However, it also clearly plays a key role in defining the distribution of stars among the varied evolutionary phases from the MSTO and beyond, thereby altering the relative distribution of stars with differing Li signatures. The fraction of FRG branch stars populating the CMD from the blue edge of the subgiant branch to the luminosity level of the RGC is controlled primarily by stellar mass. Given [Fe/H], stars above a certain mass leaving the main sequence have convective cores such that H-exhaustion occurs simultaneously within an isothermal He core; the He core mass grows with time due to H-burning in a shell outside the core. However, if the core mass fraction rises above approximately 10\%, core collapse ensues and the star evolves rapidly in \teff\ to reach the base of the vertical FRG branch, creating the Hertzsprung gap. As core contraction continues and the core temperature rises, He-ignition finally occurs under nondegenerate conditions.

For stars of lower mass, the size of the convective zone within the core declines with decreasing mass until the core is totally radiative. Under these circumstances, the fractional mass of the totally He core region grows with time, but fails to reach the critical 10\% boundary before becoming partially degenerate. The presence of degenerate gas supplies an alternative means of support which allows the star to evolve more slowly from the MSTO across the subgiant branch to the base of the vertical FRG branch. The burning H-shell slowly adds mass to the He-core, moving the star up the FRG branch until He ignites under degenerate conditions. The slower rate of evolution leads to well-populated subgiant and vertical FRG branches, as exemplified by the classic old disk cluster M67 (see, e.g. Figure 7 of \citet{TW23}). It should be noted, however, that the transition to lower mass stars not only leads to better delineation of the subgiant branch, but also weakens the fractional population of the RGC and the RG branch (FRG and AGB combined) above the luminosity of the RGC. Compare, e.g., the distribution of RG stars for NGC 2204, as exemplified by Figure 4, with a similar plot for NGC 2243 at 3.6 Gyr (Figure 4 of \citet{AT21}). Ignoring stars positioned well off the evolutionary tracks in the CMD, NGC 2204 exhibits approximately the same number ($\sim$12) of giants above (FRG and AGB) as below the luminosity of the clump. By comparison, similar counts for NGC 2243 produce a ratio of 8 (23) to 1 (3) in favor of stars below the clump. As for the actual number of stars within the clump, the approximate number is 15 for NGC 2204 as opposed to 6 in NGC 2243. Equally relevant, for NGC 2204, all but two of the stars below the RGC and half those above it have detectable Li. One potential RGC star might have detectable Li. In NGC 2243, none of the RGC or red giants brighter than the RGC have detectable Li while only half below the RGC do. Clearly the greater age of NGC 2243 dramatically alters the evolutionary distribution of the evolved stars and their Li distribution.

However, age must not be the only impactful factor, as illustrated by a direct comparison of NGC 2204 with NGC 2506, a cluster of apparently identical age (1.85 Gyr) but slightly higher [Fe/H] (-0.27) \citep{AT16, AT18b}. As shown in Figure 5 of \citet{AT18b}, the single-star FRG branch ($(B-V)$ above 0.6) of NGC 2506 contains 31 stars to the luminosity level of the RGC, all but 7 of which have measurable Li. Only 3 stars sit above the level of the clump, two of which lie well to the blue of the expected FRG branch and only have upper limits for A(Li). Approximately two dozen stars populate the RGC; only one has detectable Li. These numbers/distributions should be contrasted with the distinctly different character laid out above for NGC 2204.

In fact, NGC 2204 morphologically leans toward the younger (1.5 Gyr) but more metal-rich ([Fe/H] $\sim$ -0.1) cluster NGC 7789 \citep{BR13, BR24}. This richly populated cluster has only 6 single FRG below the level of the clump compared to 21 FRG and AGB stars above it and a dominant 56 single stars within the RGC. The fact that NGC 2204 looks morphologically younger than NGC 2506, a cluster of the same age but higher metallicity, is consistent with the pattern defined by the main sequence Li-dip. As discussed earlier, the masses bracketing the main sequence Li-dip decline by 0.04 \msun\  for each drop of 0.1 dex in [Fe/H]. The impact of this shift to lower mass is that more metal-deficient clusters exhibit the same qualitative trend with Li as metal-rich clusters, just delayed in time. For example, while NGC 2243 \citep{AT21} and M67 \citep{TW23} have almost the same age, the stars leaving the main sequence in the former cluster come from the hot side of the Li wall, while the stars populating the subgiant and giant branches in M67 came from well inside the Li-dip \citep{BO20}. NGC 2243 won't reach the Li-dip/subgiant morphology of M67 for at least 0.5 Gyr. Since the metallicity differential between NGC 2506 and NGC 2204 is less than one-fourth that of M67/NGC 2243, one expects a much smaller morphological age differential when comparing them. Therefore, if the structural parameters that generate the metallicity dependence of the mass limits for the Li-dip are reflections of a more fundamental role of metallicity/mass controlling the post-main-sequence luminosity function among first and second-ascent red giants, NGC 2204 should morphologically resemble a younger NGC 2506 at the same CMD-based isochronal age. For completeness, it should be noted that one can simply make NGC 2204 younger by adopting a larger reddening of $E(B-V)$ = 0.09. This makes the turnoff bluer, leading to higher stellar \teff\  and higher [Fe/H] by +0.03 dex. The revised age bceomes 1.7 Gyr, though the quality of the fit to the isochrones is noticeably worse.

To close the cluster comparison, we turn to the issue of MSTO rotation speeds. As demonstrated initially for 4 clusters in Figure 11 in \citet{DE19} and expanded to include NGC 2243 in Figure 11 of \citet{AT21}, the distribution of rotation speeds for stars above the wall is a strong function of age and, like the Li-dip itself, metallicity. For a younger cluster of near solar metallicity like NGC 7789, \vrot\ extends from an observational minimum near 20 \kms\ to measureable limits approaching 100 \kms. For some stars with even higher assumed \vrot\ the lines are too extended to allow plausible measurement. NGC 3680 \citep{AT09}, with a metallicity comparable to NGC 7789 but an estimated age of 1.75 Gyr, displays a much narrower range of \vrot\ with all stars located below 50 \kms. By the 2.25 Gyr age of NGC 6819 and the 3.6 Gyr age of the very metal-deficient NGC 2243, all stars but one in each cluster have \vrot\ below 25 \kms. By contrast, NGC 2506, with an age slightly greater than NGC 3680 but more metal-deficient, displays a \vrot\ spread that is dominated by stars between 25 and 65 \kms, but reaches to almost 100 \kms, much more similar to NGC 7789. Completing the pattern, Figure 8 shows that more than half the stars above $V$ = 16.5 (open circles) have no measured Li limit or detection. The stars that are plotted in the right side of Figure 8 have \vrot\ that ranges from the system limit of $\sim$20 km/s to just under 50 \kms (Table 5). The inability to measure \vrot\ for the open circles of Figure 8 stems from the same problem already noted for a subsample of stars in NGC 7789, the typical line is spread over such a wide range in wavelength that a plausible measure of its true width becomes debatable. While some of the stars without \vrot\ measures may be unresolved binaries composed of equal mass stars, especially given the almost vertical nature of the MSTO, the fraction of open circles is too large to be explained this way, especially given the analysis of similar samples in other clusters. Continuing the pattern laid out above for the post-main-sequence morphology, we conclude that the \vrot\ distribution for NGC 2204 is similar to that for NGC 2506 but, because of its lower metallicity at the same age, is more heavily weighted toward \vrot\ above 50 \kms, leaning toward a distribution more comparable to NGC 7789. Because of the lower [Fe/H], the weaker lines may become less discernible at \vrot\ above 50 \kms. As a crude test of this hypothesis use was made of the one line that remains strong, sometimes too strong, for the hot stars populating the vertical turnoff on NGC 2204: H$\alpha$. Using the interactive routine, {\it splot}, within IRAF, a Lorentzian profile was matched to the H$\alpha$ line in each spectrum. For the stars with measurable \vrot\ in Table 5, the average full-width-half-maximum (FWHM) of the line was determined to be 4.85 $\pm$ 0.40 \AA. For the stars where the line strengths were deemed too shallow to supply a plausible A(Li) measure, the comparable FWHM is 5.93 $\pm$ 0.97 \AA\ . For the extreme cases where no metal lines can be readily measured, FWHM becomes 6.62 $\pm$ 0.65 \AA.

\subsection{NGC 2204 and Galactic Li Evolution}
The metallicity of a cluster clearly impacts the evolution of Li within the cluster as a function of mass, particularly at intermediate and lower masses. Equally relevant is the manner in which cluster metallicity is tied to the initial Li abundance of the cluster. It has been recognized for decades that with a primordial A(Li) near 2.7 and a solar value at 3.3 \citep{SP12, AS09}, A(Li) must increase with time via contributions from stellar sources.  Since the mean Galactic [Fe/H] has increased from almost 0 to solar and higher over the same interval, it is generally assumed that A(Li) and [Fe/H] should be roughly correlated, a pattern confirmed by a variety of cluster and field studies (see e.g. \citet{GA20, RA20} and the discussion in \citet{AT18b}. How does NGC 2204 fit within the metallicity trend? For NGC 2506 ([Fe/H] $= -0.27$), the mean A(Li) among the stars at the vertical turnoff is 3.05 $\pm$ 0.02 (sem) from 71 stars. While some stars do scatter above the solar system value of 3.3, it is clear that the stars on the hot side of the Li-dip are systematically lower in the mean than expected for solar metallicity \citep{AT21}. At the other end of the metallicity scale, NGC 2243 ([Fe/H] $= -0.54$) has a much smaller range in $V$ for the stars on the hot side of the Li-dip, consistent with its older age. However, among this limited sample, only one star has A(Li) greater than 3.0 and the range for stars above the Li-dip extends to A(Li) $\sim$ 2.0 \citep{AT21}. Clearly NGC 2243 has a lower mean A(Li) than NGC 2506. For NGC 2204 ([Fe/H] = -0.40), the mean A(Li) for stars on the hot side of the Li-dip is 2.83 $\pm$ 0.03 (sem) from 15 stars. Keeping in mind that the sample size is modest and stars with significant rotation could not be evaluated for Li, it is still the case that no star at the turnoff of NGC 2204 outside the Li-dip has A(Li) above 3.0, and the full A(Li) range extends from 3.0 to 2.7, significantly smaller than found in NGC 2506 at the same age.

\section{Summary}
The focus of the current investigation has been the metal-deficient, moderately old open cluster NGC 2204. High dispersion spectroscopy of 167 stars, selected primarily to map the evolutionary change in Li from the MSTO to the FRG branch, has provided unusual insight into the broader question of how stellar metallicity, mass, and age can impact the observed distribution of A(Li) among evolving stars within the Galactic disk. After eliminating astrometric and radial velocity nonmembers, including unfortunately the majority of stars positioned between the MSTO and the FRG branch at the level of the clump, as well as probable binaries, EW analysis of hundreds of Fe lines in 45 evolved stars with narrow line profiles  generates a mean metallicity of [Fe/H] = -0.40 $\pm$ 0.12 (sd). Given the number of stars/lines included in the cluster average, the dominant uncertainty in the abundance remains the zero-point of the absolute [Fe/H] scale through the uncertainty in $E(B-V)$ and ultimately the \teff\ scale, though there is universal agreement that the reddening in the direction of NGC 2204 is low, consistent with the adopted value of \ebv = 0.07. The derived cluster abundance places NGC 2204 almost exactly midway between the more metal-rich NGC 2506 \citep{AT16, AT18b} and the more metal-poor NGC 2243 \citep{AT21}. 

The complications that make the MSTO abundance analyses significantly more challenging for a cluster like NGC 2204 are two-fold: (a) at lower [Fe/H], stars at a given \teff\ generally have weaker metallic lines and (b) stars of lower [Fe/H] at a given age near the MSTO are generally hotter. Added to these issues is the unexpected discovery that a significant fraction of the stars at the MSTO for NGC 2204 exhibit higher than expected rotation speeds, in many cases making any attempt at line measurement impossible.

With the basic stellar parameters in hand, spectrum synthesis was carried out for all cluster members where line broadening due to rotation was small enough to allow adequate evaluation of the impact of varying the Li line strength. This proved feasible for all the post-MSTO stars, but allowed A(Li) estimation for less than half (24 of 56) of the stars at or below the MSTO. For the red giants, the A(Li) pattern is familiar, but striking. At the base of the vertical FRG branch, A(Li) falls just below the canonically predicted Li-rich boundary for giants at 1.4. As stars evolve up the FRG branch, measurable A(Li) drops steadily to a limiting value below $-0.4$ for the star at the tip of the FRG branch. Two stars lie just beyond this tip limit, but the cool \teff\ and the spectroscopic complexity make model synthesis almost impossible. At best, one can say that these stars appear to be more Li-deficient than the hotter stars, but just how deficient is uncertain. With one exception, the stars within the RGC exhibit upper limits to A(Li) which are lower than the detectable values for the FRG branch stars at the same luminosity as the clump. If these stars evolve from the population at the red giant tip, the true upper limits for the RGC stars should be A(Li) $= -0.4$ or lower, a limit which is simply beyond the capacity of our spectra for stars at the \teff\ of the clump.  

Perhaps the most striking feature of the A(Li) CMD distribution among the evolved stars is the separation of stars above the clump into two distinct categories of detection versus upper limit, where the stars with upper limits all lie brighter/bluer than the stars with detections. With 2 exceptions that sit well above the FRG branch, the separation in color/luminosity is small, but the photometric precision is high enough to accept this separation as real. The obvious explanation for this bluer band is that they represent post-RGC stars evolving up the giant branch a second time, i.e. they are AGB stars. The spectroscopic analysis of one of these stars by \citet{CA16} supports its AGB status.

The giant population of NGC 2204 also contains an example of a persistant anomaly appearing consistently among older clusters with a sufficient population of stars, giants with rapid rotation and/or anomalous Li for their supposed position within the CMD (see \citet{AT20, SU22} for recent comprehensive discussions of this phenomenon). Since the spectrographic line signature makes the measurement of \vrot\ below $\sim$ 15 \kms\ virtually impossible, it is expected that the mean for all giants should scatter close to this value. As detailed in Section 2.2, it does. However, one star, WOCS4009, (see Figure 3) clearly has anomalously broadened lines compared to the typical effect found in most giants, implying a speed near 35 \kms. Evolved stars with \vrot\ speeds above 30 \kms\ are rare, consistently making up less than 1\%  of the K giants found in the field \citep{CA11, TA15}. The challenges to understanding these stars are multiple. First, if a star is a rapid rotator as a giant, did it start out that way on the main sequence and simply fail to spin down along the path of normal post-main sequence evolution or was there a physical mechanism which caused a normal giant to spin up? In both cases, the common solution remains interaction/mass transfer with a binary companion (see, e.g. the discussion of short-period, tidally-locked binaries in \citet{DE19}). For NGC 2204, unlike NGC 2243, the \vrot\ distribution of stars at the MSTO appears dominated by stars undergoing rapid rotation, thereby providing a ready source of potential candidates for a rapid rotator progenitor. Additionally, NGC 2204 has a modest supply of blue stragglers, commonly accepted as the products of mass transfer within a binary. Second, the evaluation of these stars is complicated by the fact that anomalous evolution likely positions them within the CMD in locations inconsistent with their actual internal state when compared to isochrones for normal single stars. Star W7017 in NGC 6819 remains a prime example \citep{AT13, CA15}. The lack of clarity regarding evolutionary phase clouds its classification as a Li-rich giant since one must first define rich relative to what. If WOCS4009 is a FRG, its A(Li) is consistent with the other stars in this evolutionary phase. If it is in the RGC, it is clearly Li-rich. This categorization is further clouded by the fact that, as a metal-deficient cluster, the primordial A(Li) for NGC 2204 appears to be closer to 2.85 than the solar metallicity reference value of 3.3. If Li is reduced by the same factor predicted for solar metallicity stars, the initial value at the base of the FRG branch should be 1.0 or less, rather than 1.4.

To close, the fundamental value of NGC 2204 remains its ability to illuminate the roles metallicity and age play within the evolution of Li for stars evolving on and after the main sequence. The trends defined and/or enhanced by past cluster studies within this series have included the shift in the position of the Li-dip with decreasing mass as [Fe/H] declines, i.e. the mass boundary of the Li wall, the narrowing of the spread in \vrot\ among stars on the hot/high mass side of the Li-wall as a cluster ages, and the correlated spread in A(Li) for the same hot/high mass stars while still on the main sequence and prior to entering the subgiant branch, again emphasizing that A(Li) does not remain constant for stars above the Li wall. With the addition of NGC 2204 to the cluster mix, a direct comparison is possible to a cluster of identical age (1.85 Gyr) but higher metallicity, NGC 2506. The contrast between the two, supposedly dominated by the difference in metallicity, reveals that the stars on the hot side of the wall are more heavily weighted toward rapid rotators than the those in NGC 2506 and the post-MSTO luminosity function shows a distribution of giants more heavily weighted toward FRG branch stars above the RGC than below, with a richer population of RGC stars, while NGC 2506 has almost no stars above the level of the RGC and the majority of this small sample has no detectable Li, possibly indicating AGB status. From the previously identified patterns, NGC 2204 more closely resembles what one expects for a younger NGC 2506, i.e. closer in appearance to the more metal-rich ([Fe/H] $\sim$ -0.1) but younger (1.5 Gyr) cluster NGC 7789. This trend is analogous to what is seen in the comparison between two clusters of much older, but similar, age ($\sim$3.6 Gyr), NGC 2243 ([Fe/H] $= -0.54$) and M67 ([Fe/H] $= +0.03$) \citep{AT21, TW23}. Stars leaving the main sequence in M67 are emerging from the cool side within the Li-dip while the stars populating the subgiant branch in NGC 2243 are still originating above the wall of the Li-dip. Thus, metallicity appears to have a broader and more nuanced impact on the distribution of stellar properties on the main sequence and beyond, over and above the simple one of setting the mass boundaries for the Li-dip, an effect which is generally downplayed or ignored in the analysis of field star distributions.

\acknowledgments
NSF support for this project was provided to BJAT and BAT through NSF grant AST-1211621, and to CPD through NSF grants 
AST-1211699 and AST-1909456. Extensive use was made of the WEBDA database maintained by E. Paunzen at the University of Vienna, Austria (http://www.univie.ac.at/webda). 

This research has made use of the ESO Science Archive Facility 
and refers to observations collected at the European Southern Observatory under ESO programme 188.B-3002(V).

Dr. Donald Lee-Brown assisted with observations and processing. The authors are grateful for the always excellent support provided by the WIYN telescope staff that made this research possible. We are also appreciative of the helpful and specific comments made by the referee which have strengthened the paper.

\facilities{WIYN: 3.5m}

\software{IRAF \citet{TODY}, MOOG \citet{SN73}, LACOSMIC \citet{VD01}, ROBOSPECT \citet{WH13}, TOPCAT citet{TOPC} }

\end{document}